\documentclass[prb,reprint,amsmath,amssymb,amsfonts,showpacs]{revtex4-1}

\usepackage{graphicx}
\usepackage{color}
\usepackage{enumerate}
\usepackage[normalem]{ulem} 
\usepackage{hyperref} 

\newcommand{\mb}[1]{\ensuremath{\mathbf{#1}}}
\newcommand{\mc}[1]{\ensuremath{\mathcal{#1}}}
\newcommand{\mr}[1]{\ensuremath{\mathrm{#1}}}

\newcommand{\im}{\ensuremath{\mathrm{Im}}}
\newcommand{\la}{\ensuremath{\langle}}
\newcommand{\ra}{\ensuremath{\rangle}}

\newcommand{\ket}[1]{\ensuremath{| #1 \rangle}}

\newcommand{\epv}[2]{\ensuremath{\langle #1 | #2 | #1 \rangle}}
\newcommand{\matel}[3]{\ensuremath{\langle #1 | #2 | #3 \rangle}}

\newcommand{\ua}{\ensuremath{{\uparrow}}}
\newcommand{\da}{\ensuremath{{\downarrow}}}

\newcommand{\Eq}[1]{Eq.~\eqref{#1}}
\newcommand{\Fig}[1]{Fig.~\ref{#1}}
\newcommand{\Sec}[1]{Sec.~\ref{#1}}
\newcommand{\App}[1]{App.~\ref{#1}}

\allowdisplaybreaks 
\begin{document}

\title{Adaptive broadening to improve spectral resolution in the numerical renormalization group}
\author{Seung-Sup B. Lee}\email{s.lee@lmu.de}
\author{Andreas Weichselbaum}\email{andreas.weichselbaum@lmu.de}
\affiliation{Physics Department, Arnold Sommerfeld Center for Theoretical Physics and Center for NanoScience, Ludwig-Maximilians-Universit\"{a}t M\"{u}nchen, Theresienstra{\ss}e 37, 80333 M\"{u}nchen, Germany}
\date{\today}
\begin{abstract}
We propose an adaptive scheme of broadening the discrete spectral data from numerical renormalization group (NRG) calculations to improve the resolution of dynamical properties at finite energies.
While the conventional scheme overbroadens narrow features at large frequency by broadening discrete weights with constant width in log-frequency,
our scheme broadens each discrete contribution individually
based on its sensitivity to a $z$-shift in the logarithmic discretization intervals.
We demonstrate that the adaptive broadening better resolves various features in non-interacting and interacting models at comparable computational cost.
The resolution enhancement is more significant for coarser discretization as typically required in multi-band calculations.
At low frequency below the energy scale of temperature, 
the discrete NRG data necessarily needs to be broadened on a linear scale.
Here we provide a method that minimizes transition artefacts in between these broadening kernels.
\end{abstract}

\maketitle

\section{Introduction}

The numerical renormalization group (NRG) is a non-perturbative method to solve quantum impurity problems~\cite{Wilson1975,Bulla2008},
with applications ranging from actual quantum impurities in mesoscopic systems to self-consistent impurity models in dynamical mean-field theory
(DMFT)~\cite{Georges1996,Bulla1999,Stadler2015}.
Owing to using a logarithmic discretization grid,
the NRG has major advantages in calculating dynamical properties.
Most importantly, it can reach arbitrarily low temperatures at comparable computational cost,
and satisfies the Friedel sum rule at temperature $T=0^+$ generally
within 1\% deviation.

The above benefits come at the cost that NRG only provides finite spectral resolution of dynamical properties at finite frequencies.
With many-body eigenstates $H \ket{E_i} = E_i \ket{E_i} $,
dynamical properties of the impurity such as the local density of states can be written in Lehmann representation as
\begin{eqnarray}
   A (\omega) = \sum_{ij} A_{ij} \delta (\omega - E_j + E_i)
\label{eq:Aspec}
\end{eqnarray}
where we use $\hbar = k_\mr{B} = 1$ in this paper, throughout.
According to the exponential coarse-graining in energy,
the conventional approach \cite{Weichselbaum2007,Bulla2001}
broadens every discrete
spectral weight $A_{ij}$ at $|\omega| > T$ with constant width $\sigma$ in
log-frequency or, equivalently, constant width-to-position ratio in
linear frequency.
As a consequence, sharp features at finite frequency
either show artificial oscillatory behavior for too small $\sigma$ or
are overbroadened otherwise. This oscillatory behavior is also inherited
by static observables e.g.~as function of temperature.
A standard prescription to deal with this situation is $z$-averaging,
which averages the discrete spectral data 
over $n_z$ logarithmic grids (shifted relative to each other by a parameter $z$), allowing the
broadening width proportional to $1/n_z$~\cite{Yoshida1990,Campo2005,Zitko2009}.
However, $z$-averaging is inherently sensitive to the precise treatment of the band edges\cite{Zitko2009} 
in that state space truncation inevitably introduces slight inequivalences for different $z$-shifts.
Eventually, this limits resolution.
Therefore besides $z$-averaging, it is desirable to have a broadening scheme that incorporates more information about the spectral data to be broadened.\cite{Freyn2009}

In this work, we propose an adaptive broadening scheme that
systematically improves resolution of dynamical properties at
finite frequencies.
For $|\omega| > T$, contrary to the conventional broadening scheme
of constant $\sigma_{(ij)}$ for all weights $A_{ij}$,
we determine the broadening width $\sigma_{ij}$ individually for each $A_{ij}$
from the sensitivity of its 
position $\omega_{ij} = E_j - E_i$ to $z$-shift, i.e.,
$\sigma_{ij} \propto \mr{d} \log | \omega_{ij} | / \mr{d} z$.
For $|\omega|<T$, the curve is further convolved with a kernel of width $\gamma < T$
on a linear frequency scale to ensure smooth behavior across $\omega=0$.
We propose a generic scheme to minimize $\gamma$
while maintaining a smooth curve for $|\omega | \lesssim T$ without artificial features at $|\omega| \sim T$.
We show that this scheme captures, for example, narrow single-particle resonances in non-interacting models
by using a clearly reduced number of $z$-shifts.
For interacting models, such as the Kondo model and the single-impurity Anderson model (SIAM),
the adaptive broadening better resolves sharp band edges,
Hubbard side peaks, or the splitting of Kondo peak by magnetic field.

This paper is organized as follows.
In \Sec{sec_NRGbasic}, we briefly review how discrete spectral data of the dynamical properties is obtained within the NRG framework.
In Sec.~\ref{sec_broad}, we present our adaptive broadening scheme.
In Sec.~\ref{sec_res}, we compare the adaptive scheme with the conventional one, by applying them to various systems at $T = 0$.
In Sec.~\ref{sec_finT}, we apply the adaptive scheme at finite $T$.

\section{Discrete data of dynamical properties by NRG}
\label{sec_NRGbasic}

\subsection{Model Hamiltonians}

The generic Hamiltonian of a quantum impurity problem can be written as
\begin{equation}
H =  
\underbrace{ 
H_\mr{imp} ( \{ d_\nu \} ) + H_\mr{cpl} ( \{ d_\nu, c_{\epsilon \nu} \} )
}_{\equiv H_0}
+ H_\mr{bath} ( \{ c_{\epsilon \nu} \} ),
\label{totHam}
\end{equation}
where $\nu$ is an index of constituent particle species (e.g.~spin, flavor, channel), $d_\nu$ is the annihilation operator at the impurity, and $c_{\epsilon \nu}$ annihilates a bath particle with energy $\epsilon$ in the bath,
satisfying $\{ c_{\epsilon \nu}, c_{\epsilon' \nu'}^\dagger \} = \delta (\epsilon - \epsilon') \delta_{\nu \nu'}$.
While different particle species interact locally within the impurity Hamiltonian
or the coupling, $H_0 \equiv H_\mr{imp} + H_\mr{cpl}$, the bath Hamiltonian is quadratic,
\begin{gather}
H_\mr{bath} = \sum_\nu \int \mr{d}\epsilon \, \epsilon \, c_{\epsilon \nu}^\dagger c_{\epsilon \nu}. \label{H_bath}
\end{gather}

In the case of the SIAM, its coupling $H_\mr{cpl}$ to the impurity is also quadratic and given by,
\begin{gather}
H_\mr{cpl}^\mr{SIAM} = \sum_\nu \int \mr{d}\epsilon \sqrt{\frac{\Gamma_\nu (\epsilon)}{\pi}} \left( d_\nu^\dagger c_{\epsilon \nu} + c_{\epsilon \nu}^\dagger d_\nu \right), \label{H_cpl}
\end{gather}
where $\Gamma_\nu (\epsilon)$ is an energy-dependent hybridization between the impurity and the bath, and $\nu \in \{\ua, \da \}$ the electronic spin.
Throughout this work we use a species-independent hybridization $\Gamma_\nu (\epsilon) = \Gamma (\epsilon) = \Gamma \Theta (D - |\epsilon|)$ and choose the half-bandwidth $D$
as unit of energy, i.e.~$D := 1$.
In the SIAM, the impurity is a single spinful electronic site with Coulomb interaction,
\begin{equation}
H_\mr{imp}^\mr{SIAM} =
    U n_{\mr{d} \ua} n_{\mr{d} \da} + \epsilon_\mr{d} n_{\mr{d}} + B S_{\mr{d},z}, 
\label{Himp:SIAM}
\end{equation}
where $n_{\mr{d}\nu} \equiv d_{\nu}^\dagger d_{\nu}$, $n_{\mr{d}} \equiv n_{\mr{d} \ua} + n_{\mr{d} \da}$ counts the number of particles on the impurity,
$S_{d,z} \equiv \tfrac{1}{2}(n_{\mr{d}\ua} - n_{\mr{d}\da})$ is the spin operator,
$U$ the Coulomb interaction strength, $\epsilon_\mr{d}$ the energy of the single-particle level,
and $B$ the Zeeman splitting due to a magnetic field in $z$-direction.
Here we consider the particle-hole symmetric case $\epsilon_\mr{d} = -U/2$.

The Kondo model is the projection of the particle-hole symmetric
SIAM onto the subspace where only one electron occupies the
impurity, in the limit $U \gg \Gamma, D$.
A Schrieffer-Wolff transformation results in
\begin{equation}
H^\mr{Kondo}_0 = J \vec{S}_\mr{d} \cdot \vec{S}_0 + B S_{d,z},\label{eq:Kondo}
\end{equation}
with $\vec{S}_0 \equiv \int_{-D}^{D} \mr{d}\epsilon \int_{-D}^{D} \mr{d}\epsilon' \sum_{\nu,\nu' = \ua,\da} c_{\epsilon \nu}^\dagger [\vec{\sigma}]_{\nu \nu'} c_{\epsilon' \nu'}$
the spin of the bath site at the location of the impurity, $\vec{\sigma}$ the Pauli spin matrices, $\vec{S}_\mr{d}$ the impurity spin operator, and $J \simeq 8 \Gamma D / \pi U > 0$.

\subsection{NRG discretization} \label{sec:NRGdisc}

A quantum impurity problem considers a localized impurity
coupled to a non-interacting bath of half-bandwidth.
In units of $D = 1$,
its continuous energies in $[-1,1]$ are discretized into logarithmic intervals split at $\pm \Lambda^{-k+1-z}$ for $k = 1,2,\ldots$
where $\Lambda > 1$ is logarithmic discretization parameter
and $z \in (0, 1]$ a discretization shift~\cite{Yoshida1990,Campo2005,Zitko2009} referred to as $z$-shift.
Here we choose $\Lambda = 2$ and $z \in \{ 1/n_z, 2/n_z, \cdots, 1 \}$.
This coarse-graining is followed by an exact mapping onto the discrete Wilson chain,\cite{Wilson1975,Bulla2008}
with exponentially decaying hopping amplitudes, i.e.~$t_n \propto \Lambda^{-n/2}$.
This introduces energy scale separation and thus justifies iterative diagonalization of the Wilson chain. 

After discretization and mapping onto the Wilson chain, the SIAM becomes
\begin{eqnarray}
H^\mr{SIAM}_N(z) &=&
\underbrace{H_\mr{imp}^\mr{SIAM} + H_\mr{cpl}^\mr{SIAM}}_{ \equiv H_\mr{0}^\mr{SIAM}} + H^\mr{bath}_N (z) ,
\label{H:SIAM}
\end{eqnarray}
now with
\begin{eqnarray}
&&H_\mr{cpl}^\mr{SIAM} =
\sum_{\nu =\ua,\da} t_{0}
 \left( d_{\nu}^\dagger f_{0\nu} + \mr{H.c.} \right) 
\label{HN:bath}
\\
&&H^\mr{bath}_N (z) =
\sum_{n = 1}^{N} \sum_{\nu =\ua,\da} t_{n}^{(z)}
   \left( f_{n-1,\nu}^\dagger f_{n\nu} + \mr{H.c} \right). 
\label{H:chain}
\end{eqnarray}
where $t_{0} = \sqrt{2\Gamma D / \pi}$ is $z$-independent,
and where $f_{n \nu}$ annihilates the electron at the chain
site $n = 0,1,2,\ldots$ with spin $\nu$.
By construction, the Kondo model maps onto a similar chain geometry,
\begin{eqnarray}
H^\mr{Kondo}_N (z) &=& H^\mr{Kondo}_0 + H^\mr{bath}_N (z),
\label{H:Kondo}
\end{eqnarray}
with $H^\mr{Kondo}_0$ as in \Eq{eq:Kondo}, but now with $\vec{S}_0 \equiv \sum_{\nu,\nu' = \ua,\da} f_{0 \nu}^\dagger [\vec{\sigma}]_{\nu \nu'} f_{0 \nu'} $.
Contrary to the original continuous Hamiltonian, 
the discrete Hamiltonians in Eqs.~\eqref{H:SIAM} and \eqref{H:Kondo} depend on $z$, due to the $z$-dependence of discretized bath $H^\mr{bath}_N$.

\subsection{Dynamical properties}

\begin{figure}
\centerline{\includegraphics[width=.49\textwidth]{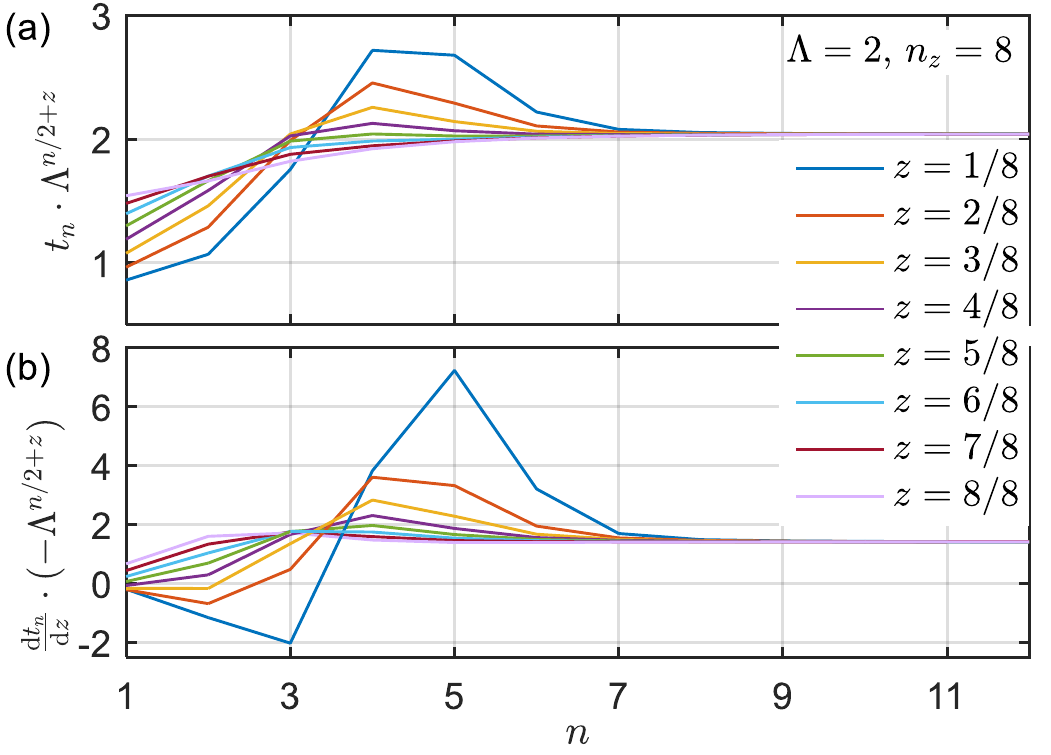}}
\caption{
(a) The hopping amplitudes $t_{n}^{(z)}$ vs. Wilson shell index $n$ and (b) their derivatives $\mr{d}t_{n} / \mr{d}z$ (obtained numerically using
$\delta z = 0.01 / n_z$), for the hybridization $\Gamma (\epsilon) = \Gamma \Theta (D - |\epsilon |)$ in units of half-bandwidth, i.e.~$D = 1$.
Since $t_{n=0}$ is independent of $z$, only data for $n>0$ is shown, which itself is independent of the value of $\Gamma$.
For large $n \gtrsim 8$, $t_{n}$ and $\mr{d} t_{n} / \mr{d}z$ follow an exponential scaling, i.e.,
$t_{n} \simeq \Lambda^{-z}  \Lambda^{(1-n)/2} (\Lambda - 1) /\log \Lambda$ and
$(\mr{d} t_n / \mr{d}z) / t_n  = \mr{d}\log(t_n^{(z)})/\mr{d}z \simeq -\log\Lambda$.
For small $n$, the values for $t_{n}$ and $\mr{d}t_{n} / \rm{d}z$ deviate from the exponential scaling.
The deviation is larger for smaller $z$, since the discretization interval at the band edge becomes narrower, i.e.~is more weakly coupled to the impurity.
}
\label{fig1_chain_coeff}
\end{figure}

Based on the argument of energy scale separation,
NRG proceeds with iterative diagonalization of the Wilson chain.
This generates a complete set of well-approximated energy eigenstates $\{ \ket{E_i^{(z)}} \}$ of the full chain.\cite{Anders2005,Anders2006}
Having the energy eigenstates $\{ \ket{E_i^{(z)}} \}$, the impurity's dynamics at finite temperature is described by local correlation functions
in the Lehmann representation\cite{Bulla2008,Anders2006,Weichselbaum2007,Weichselbaum2012:mps} [see also  \Eq{eq:Aspec}],
\begin{eqnarray}
A_z^{\mathrm{disc}}
(\omega) &=& -\frac{1}{\pi} \im \la O || O^\dagger \ra_\omega \notag \\
&=& - \frac{1}{\pi} \im \int \mr{d} t \, e^{i\omega t} \left( -i \Theta (t) \left\la [ O (t), O^\dagger ]_\pm \right\ra \right) \notag \\
&=& \sum_{ij} A_{ij}^{(z)} \, \delta ( \omega - \omega_{ij}^{(z)} ),
\label{Adisc}   
\end{eqnarray}
with 
\begin{eqnarray}
A_{ij}^{(z)} &=& |\matel{E_i^{(z)}}{O}{E_j^{(z)}}|^2
(\rho_i^{(z)} \pm \rho_j^{(z)}) \notag \\ 
\omega_{ij}^{(z)} &=& E_j^{(z)} - E_i^{(z)},
\label{lehmann:Aij}
\end{eqnarray}
where $O$ is a local operator acting at the impurity (e.g.~spin, particle creation/annihilation), $\pm$ takes $+(-)$
for a fermionic (bosonic) operator $O$,
and $\rho_i = e^{-E_i / T} / (\sum_j e^{-E_j / T})$ is the diagonal
of the density matrix at temperature $T$.
Here we employ the full-density-matrix (fdm-) NRG \cite{Weichselbaum2007,Weichselbaum2012:mps} in evaluating \Eq{Adisc}.

The dynamics of the impurity in either the SIAM 
or the Kondo model is described by the spin and frequency 
resolved $\mc{T}$-matrix for
electrons scattering off the impurity.
By using equations of motion\cite{Costi2000}, it is given by 
\begin{subequations}\label{eq:Tmatrix}
\begin{eqnarray}
\mc{T}_\nu(\omega) &=& \left\{
\begin{array}[c]{ll}
  \pi\Gamma  \la d_\nu || d_\nu^\dagger \ra_\omega & \text{ (SIAM)}\\[1ex]
  \tfrac{\pi^2}{2D} \la  O_\nu || O_\nu^\dagger \ra_\omega & \text{ (Kondo)}
\end{array}
\right.
\label{eq:Tmatrix:1}
\end{eqnarray}
For the SIAM, this leads to the impurity spectral function 
$A_\nu (\omega) = -\tfrac{1}{\pi} \im \la d_\nu || d_\nu^\dagger \ra_\omega$, whose spectral resolution 
can be improved by utilizing the impurity self-energy
$\Sigma_\nu\equiv \la [ U n_\ua n_\da, d_\nu] || d_\nu^\dagger
\ra_\omega / \la d_\nu || d_\nu^\dagger \ra_\omega$. \cite{Bulla1998}
For the Kondo model this introduces the 
local correlation function $\la  O_\nu || O_\nu^\dagger \ra_\omega$
in terms of the composite operator \cite{Costi2000}
$O_\nu \equiv [ f_{0\nu}, H^{\mr{Kondo}}_0 ] = [ f_{0\nu}, J \vec{S}_\mr{d} \cdot \vec{S}_0]$ (see \Eq{H:Kondo}).

The imaginary part of the  $\mc{T}$-matrix
defines the frequency and spin-resolved transmission probability
\begin{eqnarray}
	   T_\nu(\omega) \equiv
	  -\tfrac{1}{\pi} \im\left[\mc{T}_\nu(\omega)\right] 
\text{,}
\label{eq:Tmatrix:2}
\end{eqnarray}
\end{subequations}
which, for simplicity, will be also referred to as $T$-matrix
(note the altered font).
One has $T_\nu(\omega) \in [0,1]$,
where $T_\nu(\omega)=1$ implies perfect transmission at given 
frequency $\omega$. Furthermore, in the absence of a magnetic
field, $T_\uparrow(\omega) = T_\downarrow(\omega) = T(\omega)$, with $T(\omega)$ the spin-averaged spectral data.

\subsection{Limitations of $z$-averaging}
\label{sec:z-limits}

Within the conventional broadening scheme,\cite{Yoshida1990,Campo2005,Zitko2009,Bulla2001,Weichselbaum2007}
the spectral resolution can be improved by decreasing $\Lambda$ and by increasing $n_z$, but the improvement is limited.
First, for practical reasons, the choice of $\Lambda$ needs to be $\gtrsim 1.7$ to ensure energy scale separation, since otherwise an excessive number of states must be kept within the NRG.\cite{Weichselbaum2011}

Second, while $n_z\geq 2$ is highly attractive to gain resolution in energy space,
there is no reason to expect 
that excessive $z$-averaging,
i.e.~$n_z \to \infty$ for finite $\Lambda \gtrsim 1.7$,
can recover the exact continuum limit $\Lambda \to 1^{+}$.
Moreover, there turns out to be a practical limit in $z$-averaging, typically $n_z \lesssim 64$ [\onlinecite{Zitko2009}],
since there exist unavoidable slight inequivalences of the spectral data of different $z$-shifts.
While the coefficients $t_{n}$ in Eq.~\eqref{H:chain} scale as $\sim  \Lambda^{-n/2-z}$ for large $n$,
$z$-dependent variations of $t_{n} \Lambda^{n/2 + z}$ occur for smaller $n$ as seen in \Fig{fig1_chain_coeff}(a).
These originate from the disruption of the logarithmic scaling at the band edge:
for example, if one strictly adheres
to the discretization $\pm \Lambda^{-k+1-z}$ ($k = 1,2,\ldots$) near the band edge,
a narrow discretization interval emerges at the band edge for $z\ll 1$.
The corresponding coarse-grained level possesses a \emph{large} level energy 
since it resides at the band edge, yet is \emph{weakly} coupled to the impurity.
As a consequence, $z\to0^+$ compromises energy scale separation.
This manifests itself in a peak-like structure in the scaled
hopping amplitudes $t_{n} \Lambda^{n/2+z}$ in \Fig{fig1_chain_coeff}(a) 
that shifts towards smaller energies, i.e.~larger $n$ as $z$ is reduced.\cite{Stadler2013}
Therefore the iterative NRG diagonalization along the Wilson
chain intrinsically faces increasing difficulty with $z\to0^+$.
The resulting bias with respect to different $z$
translates into slightly uneven distribution of spectral weights
after $z$-averaging.
Though the unevenness is smoothened by large enough broadening for small $n_z$,
it introduces ``noise'' for larger $n_z$
(e.g.~see dash-dot line in Fig.~\ref{fig2_U0}(a)).
Hence the gain in spectral resolution slows down with increasing $n_z$.

\section{Broadening discrete data}
\label{sec_broad}

In order to recover the continuum from the discrete spectral data $A_z^\mr{disc}$ in Eq.~\eqref{Adisc}, we first need to gather the discrete spectral data in a suitable way.
Since we will associate each discrete weight not only with an individual energy $\omega_{ij}$ but also with an individual broadening width $\sigma_{ij}$ [specified in \Eq{sigmaij} below],
we will use a \emph{two}-dimensional binning scheme
(instead of the usual one-dimensional scheme used when all weights are broadened by the same width).
For this, we introduce a fine-grained binning in log-frequency space (about $250$ to $500$ bins per decade) as well as a linear binning in the broadening width $\sigma$
(e.g.~linearly spaced between $[0.01, \sigma_{\max}]$ with spacing $0.01$,
where $\sigma_{\max}$ can be chosen differently in different contexts;
$\sigma_{\max} = 2 \log \Lambda$ is enough at $T = 0$, while $\sigma_{\max} = 8 \log \Lambda$ was used for finite $T$ for the sake of the analysis).
A discrete spectral weight $A_{ij}$ at $\omega_{ij}$ and
broadening $\sigma_{ij}$ can then be associated with bin $k$
at frequency $\omega_k$ and broadening $\sigma_k$, i.e.~it is added to a two-dimensional array
\begin{eqnarray}
    A_{ij}^{(z)} \text{ at } (\omega_{ij},\sigma_{ij})_z \to A_{z}(\omega_k,\sigma_k).
\label{A:binning}
\end{eqnarray}
While the first dimension (binning in $\omega$) is standard within the
NRG,\cite{Bulla2008} the binning in an adaptive $\sigma$ is new.
Subsequent $z$-averaging then is easily performed 
on the level of the fine-grained binned data array in $A(\omega_k,\sigma_k)$,
\begin{eqnarray}
\bar A \equiv \langle A \rangle_z =  \frac{1}{n_z} \sum_z A_z
\label{A:zavg}
\end{eqnarray}
The broadening scheme proposed in this paper
is based on using the above $z$-averaged two-dimensional array
$\bar{A}$ as input for the following broadening formula:
\begin{eqnarray}
A(\omega) =
L_\gamma
\left[
  \sum_{k} \bar{A}(\omega_k,\sigma_k) \delta_{\bar{\sigma}_k}(\omega;\omega_{k})
\right] \text{.}
\label{A:cont}
\end{eqnarray}
Here the broadening consists of two subsequent steps:
First the discrete spectral data in bin $\bar{A}(\omega_k,\sigma_k)$ is broadened using 
a standard NRG log-Gaussian broadening kernel\cite{Weichselbaum2007,Bulla2008} $\delta_{\bar{\sigma}_k}(\omega;\omega_{k})$ [see \Eq{logGauss} below],
centered around $\omega_{k}$,
yet with individual broadening width
$\bar{\sigma}_{k} \equiv \tfrac{\alpha}{n_z}\sigma_{k}$,
where $\alpha$ is an overall constant prefactor
[see \Eq{sigma:k} below].
This first step is applied to the spectral data at \emph{all}
frequencies, yet it only generates smooth data for frequencies $\omega>T$.
It still leaves pronounced discrete features for $|\omega| \lesssim T$.

In a second step we employ uniform linear broadening $L_\gamma[\cdot]$
of optimized width $\gamma \lesssim T$ [cf.~\Eq{delta:lin} below].
The latter is again applied to the full frequency range,
thus there is no need to specify a transition function
for switching from log-Gaussian to linear broadening
(in contrast to the conventional scheme of Ref.~\onlinecite{Weichselbaum2007}).
Consequently, while there is minor numerical overhead involved
by considering the full frequency range, the latter minimizes
artefacts in the final spectral data for intermediate frequencies
$|\omega| \sim \gamma$. For exponentially large frequencies
$|\omega| \gg T \gtrsim \gamma$, 
where the data obtained from the first log-Gaussian broadening step is already smooth,
the uniform broadening $L_\gamma[\cdot]$ has
negligible effect, so eventually may be skipped there.

Another approach to improve the spectral resolution\cite{Osolin2013}
first converts the discrete data to the imaginary-frequency domain via the Hilbert transform and 
then applies the analytic continuation to obtain real-frequency curves (without imposing broadening).
However, the analytic continuation is numerically ill-posed and thus subject to error.
Indeed, as the impurity solver of the DMFT, the NRG is advantageous exactly for the reason that 
the dynamical properties are directly computed on the real-frequency axis without the analytic continuation.\cite{Stadler2015}

In the following, we discuss the individual steps above in more detail.
For simplicity, we start with the conventional and adaptive broadening schemes for the regime $|\omega| > T$, in which all the weights are logarithmically distributed.
Then we address the regime $|\omega| \lesssim T$.
For all of the following discussion we already assume $z$-averaged discrete NRG data, unless indicated otherwise.

\subsection{Broadening for $|\omega| > T$}
\label{sec_log_broad}

\begin{figure}
\centerline{\includegraphics[width=.49\textwidth]{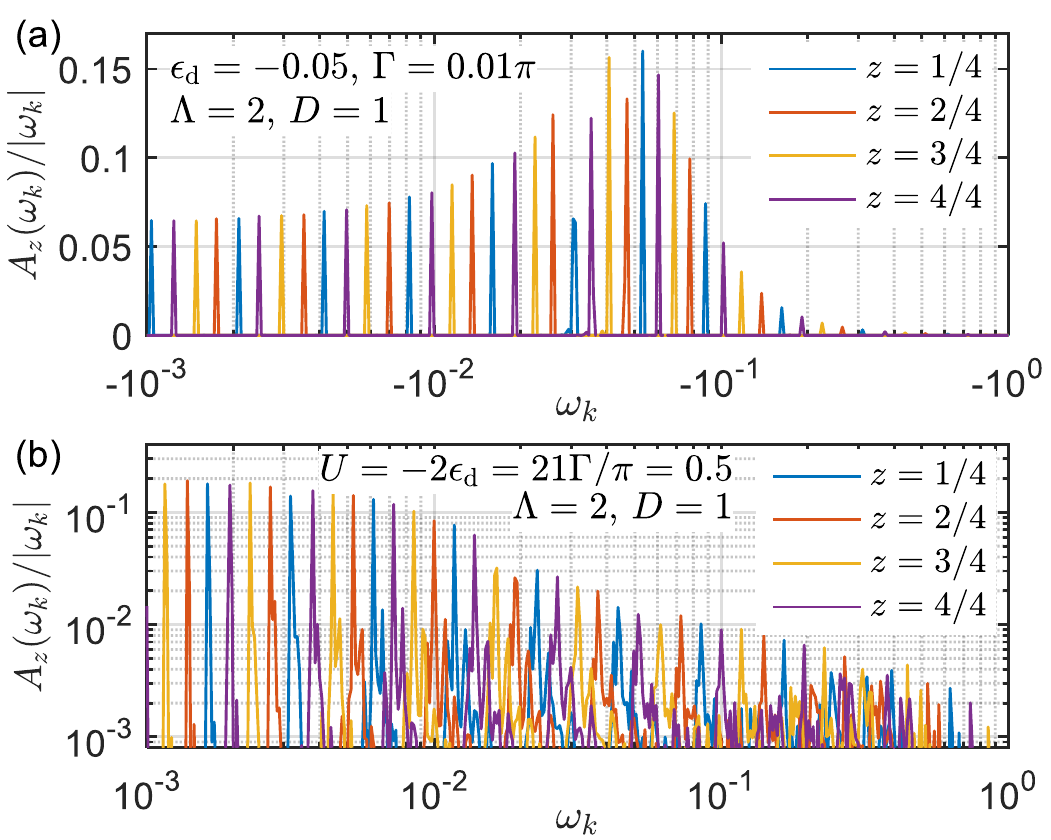}}
\caption{
Discrete spectral data $A_z (\omega_k) \equiv \sum_{\sigma_k} A_z (\omega_k, \sigma_k)$ for (a) the non-interacting resonant level model [see \Eq{H_NRLM}] and (b) the SIAM at $T=0$.
Different colors denote different $z$-shifts.
Since the bin $k$ at frequency $\omega_k$ has width
proportional to $|\omega_k|$ in the frequency domain,
$A_z (\omega_k) / |\omega_k|$ is related with the height of a subsequently broadened curve.
The discrete spectral weights are bunched, and most of bunches are uniformly spaced in log-frequency space with the spacing $\Delta (\log |\omega|) \simeq (\log \Lambda) / n_z$ after $z$-averaging.
But in certain regimes [$\omega_k \sim \epsilon_\mr{d} = -0.05$ for (a) and $\omega_k \sim U/2 = 0.25$, $\omega_k \sim T_\mr{K} \simeq 0.01$ for (b)], bunches are widely spread or irregularly distributed:
the adaptive broadening scheme systematically applies the narrower broadening width to these bunches,
leading to the improved spectral resolution in such regimes (see Fig.~\ref{fig2_U0} and \ref{fig3_B0_Anderson}).
See Secs.~\ref{sec_U0} and \ref{sec_B0} for detail.
}
\label{fig8_bunch}
\end{figure}

The discrete spectral contributions of $A_z^{\mathrm{disc}} (\omega)$ are bunched in log-frequency space,
where average distance between bunches $\Delta (\log |\omega|) \simeq (\log \Lambda) / n_z$ after $z$-averaging,
as illustrated in \Fig{fig8_bunch}.
This originates from the underlying discretization grid:
due to the logarithmic discretization, the discretized energy levels of the bath are located at $\sim \pm \Lambda^{-k+1-z}$ ($k = 1,2,\ldots$),
that is, uniformly spaced in log-frequency space (except near the band edges).
By coupling the impurity to the bath,
the energy levels of the total system are shifted from the bare bath energy levels,
but the overall logarithmic distribution remains.

We use the standard NRG log-Gaussian broadening kernel\cite{Bulla2008,Weichselbaum2007}
\begin{subequations} 
\begin{eqnarray}
\delta_{\bar{\sigma}_k}(\omega;\omega_k)
&=& \frac{\Theta(\omega \omega_k)}{\sqrt{\pi} \bar{\sigma}_k | \omega_k |}
e^{-\left( \frac{\log | \omega / \omega_k|}{\bar{\sigma}_k} -
\frac{\bar{\sigma}_k}{4} \right)^2 },
\label{logGauss}
\end{eqnarray}
which preserves the spectral sum rule and Kondo peak height, \cite{Weichselbaum2007}
yet where we explicitly introduce an individually determined broadening width
\begin{eqnarray}
\bar{\sigma}_{k} &\equiv& \underbrace{\tfrac{\alpha}{n_z}}_{\equiv \alpha_z} \sigma_{k}
\label{sigma:k}
\end{eqnarray}%
\end{subequations}%
with $\alpha$ an overall constant prefactor of order 1.
In this work, we choose $\alpha\in[1,2]$ and specify its value
with each figure below.
We use $\alpha=1$ for the non-interacting
resonant level model, $\alpha \in \{1.5, 2\}$ for the SIAM and Kondo
model at $\Lambda=2$, and $\alpha = 2$ also
for extremely large $\Lambda\lesssim16$ [cf.~\Fig{fig10_largeL}].
The bar on the l.h.s.~of \Eq{sigma:k} is a reminder that this is the final broadening used on the $z$-averaged data as in \Eq{A:zavg}.

Conventional broadening schemes \cite{Bulla2001,Weichselbaum2007}
use a constant  $\sigma_k = \log\Lambda$ for all discrete spectral weights
(i.e.~$\bar{\sigma}_{(k)} = \mathrm{const} \lesssim \log\Lambda$).
This choice is natural for the discrete weights deep inside fixed-point regimes such as the stable low-energy fixed-point 
where the spectral data is featureless and the spectral data appears bunched at distance $\Delta (\log |\omega|) = (\log \Lambda) / n_z$
in log-frequency space, suggesting $\alpha=1$.
However, this leads to overbroadening of sharp spectral features at 
finite frequencies $|\omega|\gtrsim T$
where discrete weights are distributed more irregularly
in relation to the underlying physics.

\subsubsection{Adaptive broadening $\sigma_k$}

The broadening scheme proposed in this paper uses the log-Gaussian in \Eq{logGauss} with the adaptive broadening width in \Eq{sigma:k} where
\begin{equation}
\sigma_{ij} \xrightarrow{\text{binning}} \sigma_k.
\label{sigma:adapt}
\end{equation}
Here $\sigma_{ij}$ is determined for each spectral weight $A_{ij}$ at frequency $\omega_{ij}$ for an arbitrary but fixed $z$-shift as in \Eq{Adisc},
and then binned according to \Eq{A:binning}.
In the following we describe and motivate a scheme for computing
$\sigma_{ij}$ for an elementary spectral contribution $A_{ij}$
prior to $z$-averaging.

For the sake of the argument, suppose the discrete data $A_z^\mathrm{disc} (\omega)$ can be obtained exactly by solving $H(z)$ in \Eq{H:SIAM} without any truncation.
As the coefficients in $H(z)$ are continuous functions of $z$,
the frequency $\omega_{ij}^{(z)}$ associated with each discrete spectral weight $A_{ij}^{(z)}$
shifts as a function of $z$.
In particular, a shift $z \to z+1$ shifts the discrete data onto itself
(except for the very band edge).
Now to fill the distance between a delta-peak of weight $A_{ij}^{(z)}$ at
frequency $\omega_{ij}^{(z)}$ and another delta-peak of weight $A_{ij}^{(z + \Delta z)}$ at $\omega_{ij}^{(z + \Delta z)}$,
with $\Delta z \equiv 1/n_z \ll 1$,
a sensible choice for the broadening width $\sigma_{ij}$ for $A_{ij}^{(z)}$ is to use the resulting shift in log-frequency scale,
$\left| \log |\omega_{ij} ^{(z)}| - \log |\omega_{ij}^{(z + \Delta z)}| \right|$,
up to a factor $\sigma_0$ of order 1.
To leading order we thus choose
\begin{subequations} \label{sigmaij}
\begin{eqnarray}
\sigma_{ij}^{(z)} &\equiv& \bigl | \tfrac{\mr{d} \log |\omega_{ij} ^{(z)} |}{\mr{d} z} \bigr | 
= \tfrac{1}{| \omega_{ij} ^{(z)} |} \bigl | \tfrac{\mr{d} E_j}{\mr{d} z} - \tfrac{\mr{d} E_i}{\mr{d} z} \bigr | \\
&=& \tfrac{1}{| \omega_{ij} ^{(z)} |}
\bigl | \epv{E_j}{\tfrac{\mr{d} H}{\mr{d} z}} - \epv{E_i}{\tfrac{\mr{d} H}{\mr{d} z}} \bigr |
\text{,}
\end{eqnarray}  
\end{subequations}
where the last expression is evaluated within an NRG run at fixed $z$-shift.
Here we used the Hellmann-Feynmann theorem $\tfrac{\mr{d}E_{i}}{\mr{d}z} = \epv{E_{i}}{\tfrac{\mr{d}H}{\mr{d}z}}$,
given that the NRG is dealing with (approximate) eigenstates $|E\rangle_i$ of the Hamiltonian $H$
(for further details, see \App{app_deg}).
Therefore overall the adaptive broadening $\sigma_{ij}^{(z)}$ is computed as the lowest-order response to the perturbation 
$\tfrac{\mr{d}H}{\mr{d}z}$. [Actually, the full perturbation is
$\tfrac{\mr{d}H}{\mr{d}z} \cdot \Delta z$; the factor $\Delta z = 1/n_z$, however, has already been
split off in \Eq{sigma:k}; hence $\tfrac{\mr{d}H}{\mr{d}z}$ will
be referred to as the perturbation here.]

Finally, the perturbation $\tfrac{\mr{d}H}{\mr{d}z}$ in \Eq{sigmaij}
for the Wilson chain for a given $z$-shift is obtained numerically as follows:
the standard Lanzcos tridiagonalization of the bath is performed for two closeby shifts $z$ and $z+\delta z$,
with e.g.~$\delta z=0.01 / n_z \ll  \Delta z$.
This gives two Wilson chains with slightly altered coefficients $t_{n}$.
The perturbation $\tfrac{\mr{d}H}{\mr{d}z}$ is
therefore defined in the same geometry as the Wilson chain,
but with different hopping amplitudes
\begin{equation}
\tfrac{\mr{d}t_{n}}{\mr{d}z } \simeq \tfrac{1}{\delta z}
\bigl( t_{n}^{(z+\delta z)} - t_{n}^{(z)} \bigr) ,
\label{eq:deltat}
\end{equation}
instead of $t_{n}$.
A typical set of the coefficients $\mr{d} t_{n } / \mr{d}z $ of the perturbation $\mr{d}H/\mr{d}z$ is presented in \Fig{fig1_chain_coeff}(b).
For simplicity, we use the particle-hole symmetric hybridization
$\Gamma (\epsilon) = \Gamma (-\epsilon)$ throughout this work.
Hence the onsite energy for each Wilson site is zero
by symmetry and thus trivially independent of $z$-shifts.
For general $\Gamma (\epsilon)$, nevertheless, the onsite energies can be simply incorporated within the tridiagonalization,
with the numerical derivatives of the resulting onsite
energies along the Wilson chain computed analogously to \Eq{eq:deltat}.
See also \App{app_LD} for further details.

Note that the two Lanzcos tridiagonalizations for slightly different $z$-shifts above effectively address a \emph{different} one-particle basis $f_{n\nu}$ [cf.~\Eq{H:chain}] via a slightly shifted coarse-graining of the bath.
However, this leads to two Wilson chains of identical structure, differing only in their parameters.
Hence without restricting the above argument, the operators $f_{n\nu}$ may be considered independent of $z$-shifts,
and the only changes in the bath are described by the coefficients $t_{n}^{(z)}$.
With this, the diagonal matrix elements of the perturbation, $\epv{E_i}{\tfrac{\mr{d}H}{\mr{d}z}}$ in \Eq{sigmaij}, can be straightforwardly evaluated directly during the iterative diagonalization of NRG.

\subsection{Broadening for $|\omega| \lesssim T$
\label{sec:Tfinite}}

Finite temperature $T$ introduces an energy scale where energy scale separation necessarily comes to a halt.
Even if the Wilson chain itself is semi-infinite,
finite $T$ introduces an effective finite length of the chain $n_T \sim -2 \log_\Lambda T$ within fdm-NRG.\cite{Weichselbaum2007,Weichselbaum2012:mps}
This intrinsically limits the energy resolution to the order of $T$.
Consequently, the log-Gaussian broadening must eventually
be replaced by a linear broadening scheme for $|\omega| \ll T$.
This also ensures that the spectral data for positive and negative frequencies
is smoothly connected across $\omega=0$.
In practice, this is achieved by using various linear broadening kernels, 
such as Gaussians \cite{Weichselbaum2007} or Lorentzians \cite{Bulla2001}
of width $\gamma\lesssim T$ in linear frequency.

However, the transition from the log-Gaussian to linear broadening
involves some arbitrariness, and tends to
introduce artificial features at intermediate frequencies at $|\omega| \sim T$.
While these artefacts can be reduced to some extent by carefully
combining two kernels \cite{Weichselbaum2007},
it is difficult to have a generic scheme to avoid artefacts for a general parameter regime where pronounced spectral features occur on the scale of temperature.
This is specifically relevant for DMFT type calculations. \cite{Georges1996,Bulla1999,Bulla2008,Bulla2001,Stadler2015}

In many cases, discretization artefacts can be systematically
suppressed at all frequencies, including both $|\omega| \lesssim T$
as well as $|\omega| \gg T$, by exploiting self-energy improved
spectral functions, \cite{Bulla1998} to be referred to as ``$\Sigma$-improved'' in the analysis below.
However, as this self-energy ``trick'' is the post-processing of smoothened spectral data, the main focus here is to 
find an optimal way of \emph{directly} broadening discrete spectral data before any post-processing.

The essential observation here is that we would like
to avoid altogether a transition function from one broadening kernel
to another, i.e.~log-Gaussian to linear broadening.
To achieve this, we perform the log-Gaussian broadening for \emph{all} frequencies, 
followed by a linear broadening that again also operates on \emph{all} frequencies,
i.e.~we use a convolution of two broadening kernels as in \Eq{A:cont}.
This scheme thus avoids constructing an ad-hoc scheme for transitioning from one type of broadening to another.

The order in which the broadening kernels are applied is important.
Since the log-Gaussian broadening does not affect the function at $\omega = 0$, by construction,
the value of $T$-matrix at $\omega = 0$ is determined at the stage of the linear broadening.
If one applies the linear broadening first,
the value $T(0)$ is determined directly from discrete data, so is subject to numerical noises.
On the other hand, if the log-Gaussian broadening is applied first,
the linear broadening acts on the curve which is already smooth for $|\omega| \gtrsim T$ in log-frequency space [see blue line in \Fig{fig6_finT}(a)],
so results in smooth curves across $\omega = 0$.
Solid and dashed lines in \Fig{fig6_finT}(c,d) illustrates the results from the ``correct'' and opposite orders, respectively:
with the opposite broadening order, the value $T(0)$ is clearly shifted from the correct value $\simeq 1$
and the fitting error $\varepsilon_\mr{RMS}$ is much larger
(see Sec.~\ref{sec_gamma} for detail).

For the linear broadening using the convolution
$L_\gamma [ g ] = \int \mr{d}\omega' \delta_\gamma (\omega - \omega') g(\omega')$
in \Eq{A:cont},
we choose the derivative of the Fermi-Dirac distribution
function $f(\omega)$,\begin{equation}
\begin{aligned}
\delta_\gamma(\omega - \omega')  &\equiv -\tfrac{\mr{d}}{\mr{d}\omega} f(\omega - \omega') \\
&= \tfrac{1}{2 \gamma} \left( 1 + \cosh \tfrac{\omega - \omega'}{\gamma} \right)^{-1} ,
\end{aligned}
\label{delta:lin}
\end{equation}
at an effectively reduced ``temperature'' $\gamma < T$.
This kernel decays more slowly for large frequencies than a regular Gaussian.
Hence it promises somewhat smoother data, whereas it still decays exponentially, in contrast to Lorentzians \cite{Bulla2001}.
Last but not least, the above choice is also motivated by
the empirical fact that the linear conductance, which exactly
corresponds to a spectral function convolved with 
$-\mr{d}f/\mr{d}\omega |_{\gamma = T}$ at $\omega = 0$, can be very accurately
computed within the NRG by convolving
the \emph{discrete} spectral
data of the NRG with $-\mr{d}f/\mr{d}\omega$.\cite{Weichselbaum2007}

It remains to find the optimal value for the linear broadening width $\gamma$.
It should be chosen such that 
(i) it removes the residual discrete features from the prior log-Gaussian broadening for frequencies $|\omega| \lesssim T$,
yet (ii) that it minimally overbroadens the remaining data otherwise.
Large $\gamma$ guarantees smooth data as argued above,
which in practice means that $\gamma=T$ is more than sufficient
to obtain smooth data. Minimizing $0<\gamma<T$ in an
systematic fashion while maintaining smooth data to
within about 0.2\% variations will be addressed in Sec.~\ref{sec_finT} below.

\section{Results at $T = 0$}
\label{sec_res}

In this Section, we compute the $T$-matrix in \Eq{eq:Tmatrix} for non-interacting and interacting impurity models at $T=0$,
and demonstrate that the adaptive broadening scheme provides overall better spectral resolution than the conventional scheme.

\begin{figure}
\centerline{\includegraphics[width=.49\textwidth]{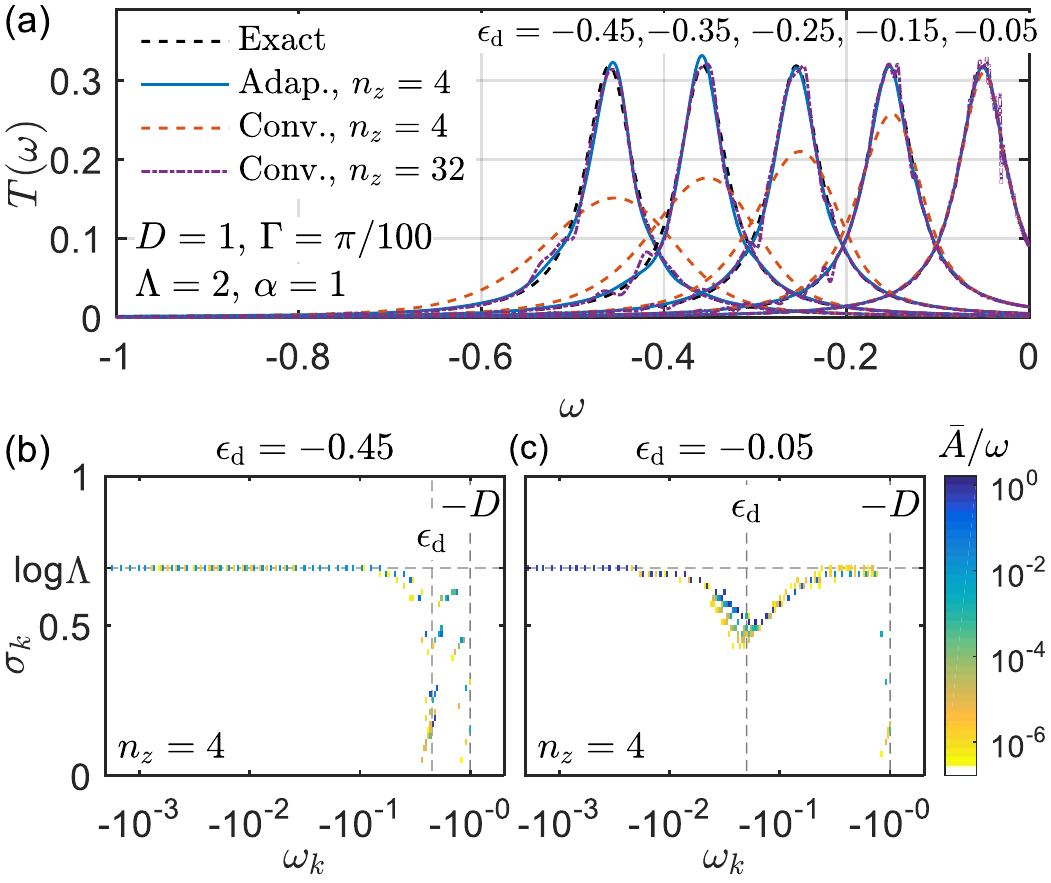}}
\caption{
Adaptive broadening for the $T$-matrix in 
the spinless non-interacting resonant level model.
Panel (a) shows the exact $T (\omega)$ [Eq.~\eqref{Aw_RLM}, black dashed line] and the NRG results with the conventional and adaptive broadening schemes for different $\epsilon_\mr{d}$.
For small $n_z$, the conventional scheme (strongly) overbroadens the resonance peaks (red dashed line), 
while the adaptive scheme (blue solid line) nicely reproduces the exact $T(\omega)$.
In contrast, the conventional broadening scheme can capture the sharp peak shapes only for much larger $n_z$ (purple dash-dot line), 
but also acquires noise [cf.~\Sec{sec:z-limits}].
Panels (b) and (c) show the binned discrete data $\bar{A} (\omega_k, \sigma_k) / |\omega_k|$ ($\bar{A}/\omega$ for short; see text) for $\epsilon_\mr{d} = -0.45$ and $\epsilon_\mr{d} = -0.05$, respectively.
The weights near $\omega_k \simeq \epsilon_\mr{d}$ as well as near the band edge $-D$ have clearly reduced adaptive broadening width $\sigma_k < \log \Lambda$.
Thus the adaptive broadening can capture the sharp features near $\omega_k \simeq  \epsilon_\mr{d}$ and band edges $| \omega_k | = D$ (not shown in (a))
much better than the conventional broadening scheme
for fixed $n_z$.
}
\label{fig2_U0}
\end{figure}

\subsection{Noninteracting resonant level model}
\label{sec_U0}

We apply the adaptive broadening scheme to the noninteracting resonant level model of spinless fermions,
\begin{equation}
H = \epsilon_\mr{d} d^\dagger d + \int\limits_{-D}^{+D} \mr{d}\epsilon 
\left(
   \sqrt{\tfrac{\Gamma}{\pi}} \left( d^\dagger c_{\epsilon } + c_{\epsilon}^\dagger d \right)
+ \epsilon \, c_{\epsilon}^\dagger c_{\epsilon}
\right),
\label{H_NRLM}
\end{equation}
which is the spinless one-band SIAM with $U = 0$.
Since the Hamiltonian is quadratic, the $T$-matrix for particles scattering off
the impurity $T (\omega) = \pi \Gamma A(\omega) = - \Gamma \, \im \la d || d^\dagger \ra_\omega$ has the closed form
\begin{eqnarray}
T(\omega) = \frac{\Gamma^2 \, \Theta \left( D - |\omega| \right)}
{\left( \omega - \epsilon_\mr{d} - \frac{\Gamma}{\pi} \log \left| \frac{D+\omega}{D-\omega} \right| \right)^2 + \Gamma^2},
\label{Aw_RLM}
\end{eqnarray}
with $D=1$ as usual.
When $\Gamma, |\epsilon_\mr{d}| \ll 1$,
$T(\omega)$ approximates a Lorentzian centered at $\epsilon_\mr{d}$
of width $2\Gamma$.
For the simulation of this model using NRG, 
we kept up to 200 states in each step of iterative diagonalization.

Though the model is quadratic and simple, the NRG with the conventional broadening is very inefficient in reproducing $T(\omega)$
in Eq.~\eqref{Aw_RLM} when the resonance is sharp in the sense $\Gamma \ll |\epsilon_\mr{d}|$.
Since every discrete weight is broadened with the fixed width-to-position ratio,
the resonance in $T(\omega)$ cannot be narrower
than $|\epsilon_\mr{d}| \cdot \tfrac{\alpha}{n_z} \log \Lambda$.
As seen in Fig.~\ref{fig2_U0}(a), this makes the peaks overbroadened when $\Gamma/|\epsilon_\mr{d}| \ll \log \Lambda / n_z$ as compared to the exact $T(\omega)$ [cf.~\Eq{Aw_RLM}].
Also, Fig.~\ref{fig2_U0}(a) shows that excessive $z$-averaging with $n_z = 32$ does not only sharpen the resonance peak but also introduces noise as discussed in \Sec{sec:z-limits}.
In contrast, the adaptive broadening already captures the resonance peaks at much lower $z$-averaging ($n_z=4$).
This can be understood by analyzing the distribution of discrete binned spectral weights $\bar{A}(\omega_k,\sigma_k)$
which are $z$-averaged as in \Eq{A:zavg}.

Figure~\ref{fig2_U0}(b,c) presents a snapshot of the
adaptively determined broadening by directly plotting the
binned data $\bar{A}(\omega_k,\sigma_k) / |\omega_k|$,
which is related with the height of a subsequently broadened curve
as mentioned in \Fig{fig8_bunch}.
Hence for the purpose of analysis of the discrete data later in this paper,
we show the array $\bar{A}(\omega_k,\sigma_k) / |\omega_k|$ ($\bar{A}/\omega$ in short).
In \Fig{fig2_U0}(b,c) then,
three features can be distinguished in the distribution of $\bar{A}/\omega$:
(i) For $|\omega_k - \epsilon_\mr{d}| \gg \Gamma $,
the weights are concentrated along the line $\sigma_k \simeq \log \Lambda$.
These weights, including the stable low-energy fixed point regime, are 
almost uniformly distributed in equally spaced bunches in log frequency.
Consequently, the adaptive broadening assigns the same broadening width as the conventional one;
$\bar{\sigma}_k = \tfrac{\alpha}{n_z} \log \Lambda$.
(ii) At the maximum of $T(\omega)$ 
around $\omega_k \simeq \epsilon_\mr{d}$, the weights in $\bar{A}/\omega$
show a clearly reduced broadening width $\sigma_k$
that can be much narrower
than in the conventional scheme [e.g. see panel (b)]. This is important
in order to capture the sharp peak structure.
(iii) Near the band edges $|\omega_k| \simeq D = 1$,
similar to (ii), again the weights have clearly reduced $\sigma_k < \log \Lambda$.
Here they describe the sharp edges of $A (\omega)$ at $\omega = \pm D$ in \Eq{Aw_RLM},
which originates from the sharp edge of the hybridization
function $\Gamma (\epsilon)$.

\begin{figure}
\centerline{\includegraphics[width=.49\textwidth]{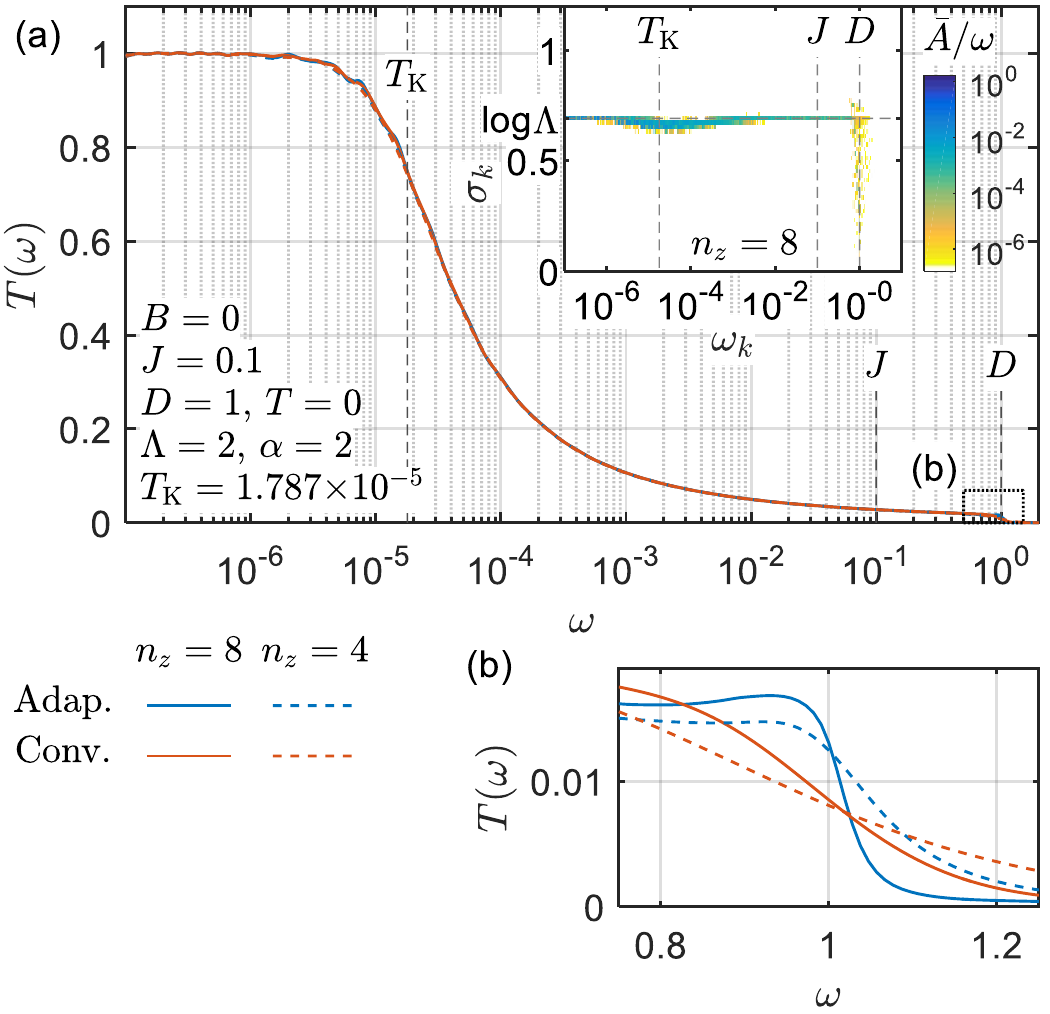}}
\caption{
$T$-matrix of the Kondo model for $B=T=0$. 
Due to total spin and particle-hole symmetries, 
$T_\nu (\pm \omega) = T (\omega)$.
(a) Broadened curves with different broadening schemes (adaptive or conventional) and number of $z$-shifts ($n_z=4$ and 8).
(Inset) Discrete spectral data $\bar{A}/\omega$ for the data in the main panel.
(b) Close-up near the band edge $D = 1$.
The adaptive broadening better resolves the sharp edge at $| \omega | = D$,
which stems from the reduced broadening
$\sigma_k < \log \Lambda$  at $|\omega_k| \simeq D$ in the inset of (a).}
\label{fig3_B0_Kondo}
\end{figure}

\begin{figure}
\centerline{\includegraphics[width=.49\textwidth]{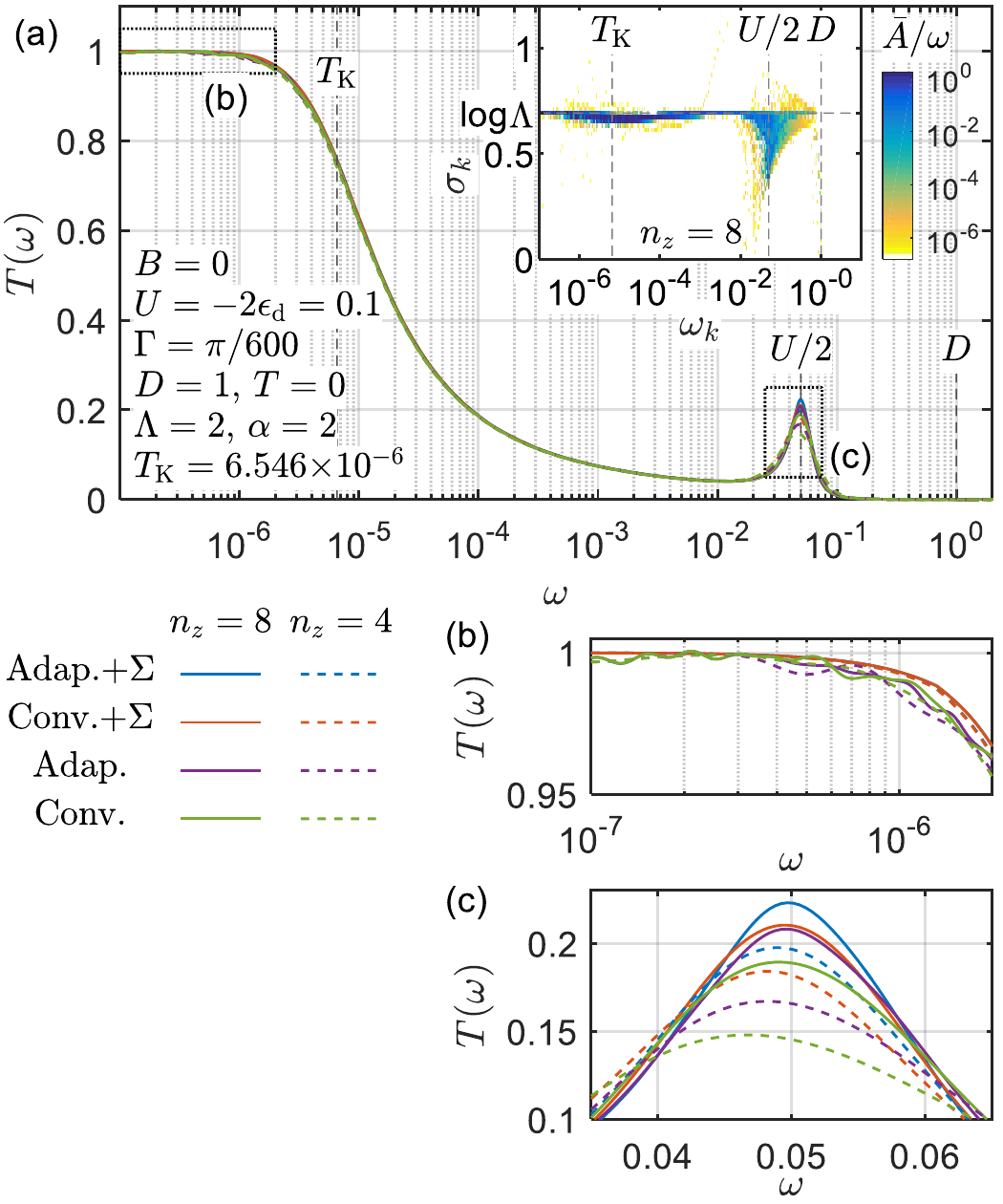}}
\caption{
$T$-matrix of the SIAM for $B=T=0$.
Due to total spin and particle-hole symmetries, 
$T_\nu (\pm \omega) = T (\omega)$.
(a) Broadened curves with different broadening schemes (adaptive or conventional), number of $z$-shifts ($n_z = 4$ and 8),
and partly improved by self-energy ($\Sigma$).
(b) Close-up of the Kondo peak at $\omega \ll T_\mr{K}$, showing convergence to within 1\%.
Here the blue solid (dashed) line coincides with the red solid (dashed) line.
(c) Close-up of the Hubbard side peak centered at $\omega = U + \epsilon_\mr{d}= U/2$.
The self-energy removes discretization related noise at $\omega \ll T_\mr{K}$ and sharpens the Hubbard side peaks.
The inset to panel (a) presents the discrete spectral data $\bar{A}/\omega$ for the data in the main panel.
It shows that the adaptive broadening reduces the broadening $\sigma_k$ for frequencies $\omega_k \simeq U/2$, and hence enhances
the resolution of the side peak.
This is more significant for smaller $n_z$ or without self-energy.}
\label{fig3_B0_Anderson}
\end{figure}

\subsection{Kondo and SIAM at $B = 0$}
\label{sec_B0}

In this Section and the next, we analyze the dynamical
behavior of the Kondo model and the SIAM over a wide range
of magnetic field $B$, starting with the case $B = 0$.
Here the frequency resolved $T$-matrix [cf.~\Eq{eq:Tmatrix}]
shows the well-known Kondo resonance at $\omega = 0$
as seen in Figs.~\ref{fig3_B0_Kondo} and \ref{fig3_B0_Anderson}
for the Kondo model and the SIAM, respectively.
The height of the Kondo peak is determined by Friedel sum rule.
The NRG gives $T(\omega = 0)\simeq 1$
accurately to within 1\% error in either model.
When also self-energy is exploited for improved spectra data
in the SIAM, the error further decreases.
The width of the Kondo peak is the Kondo temperature
$T_\mr{K}$, up to an $O(1)$ factor
depending on the precise definition.
For both Kondo and Anderson models, we determined $T_\mr{K}$ as the frequency $\omega$ at which the dynamical impurity spin susceptibility $\chi_s (\omega)
\equiv - \tfrac{1}{3 \pi} \im \la \vec{S}_\mr{d} || \vec{S}_\mr{d} \ra_\omega $
becomes maximum~\cite{Hanl2014} (here the factor $1/3$ comes from the average
of the components in $\vec{S}_\mr{d} \cdot \vec{S}_\mr{d}$
when exploiting SU(2) spin symmetry, e.g.~see Ref.~\onlinecite{Weichselbaum2012:sym}).
In the simulations, we kept up to 500 multiplets 
in each step of the iterative diagonalization when exploiting $\mr{SU}(2)_{\mr{spin}} \otimes \mr{SU}(2)_{\mr{ph}}$ for spin and particle-hole symmetry, respectively.

Both the Kondo as well as the SIAM share the same Kondo physics around $|\omega| \lesssim T_\mr{K}$,
as well as the sharp cutoff outside the band edge
$|\omega| >D $. The Hubbard side peaks of the SIAM, however,
are absent in the Kondo model.
The adaptive broadening enhances the resolution of the high-frequency
features, as in Fig.~\ref{fig3_B0_Kondo}(b) and Fig.~\ref{fig3_B0_Anderson}(c).
Quite generally, the enhancement is more significant 
where features are overbroadened in the conventional scheme.
In particular, this is the case when $n_z$ is smaller,
$\Lambda$ is larger, and in case of the SIAM, no self-energy is used.
In \Sec{sec_largeL}, we further discuss the performance of adaptive broadening for large $\Lambda$.
Here in the insets of Fig.~\ref{fig3_B0_Kondo}(a) and
Fig.~\ref{fig3_B0_Anderson}(a),
we show for fixed $\Lambda=2$
where and how the adaptive broadening of the discrete spectral data $\bar{A}/\omega$
enhances the resolution.
The broadening is clearly reduced at the band edges $|\omega_k| = D$,
and for the SIAM also around the Hubbard side peaks,
$|\omega_k| \simeq U/2$, in that
the distribution in $\sigma_k$ clearly spreads to lower values.
Also, for the SIAM, the spread in $\sigma_k$ at $|\omega_k| = D$
is much less pronounced, since at $|\omega_k| \gg U/2$
the spectral weight decays more strongly as compared to the Kondo model.

The adaptive broadening gives quantitatively the same Kondo peak shape as the conventional broadening, 
since it is located around $\omega=0$ where the resolution is exponentially refined due to the underlying logarithmic discretization.
Furthermore, given the relatively slow logarithmic corrections that enter the Kondo peak shape,
the discrete weights $\bar{A}/\omega$ for $|\omega_k| \lesssim T_\mr{K}$ are mostly distributed around $\sigma_k \simeq \log \Lambda$.

\begin{figure}
\centerline{\includegraphics[width=.49\textwidth]{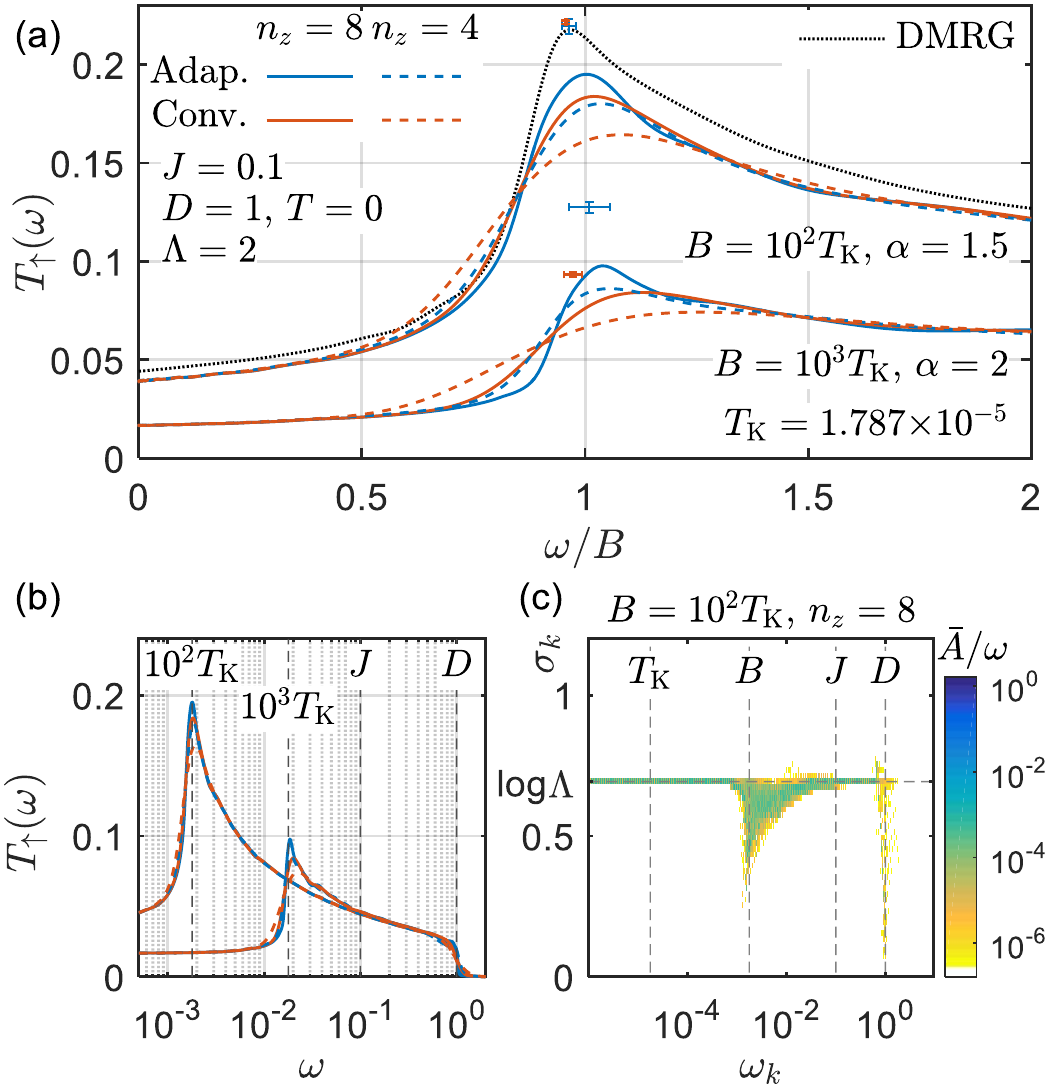}}
\caption{
$T$-matrix of the Kondo model for large magnetic field $B \gg T_\mr{K}$ at $T=0$.
Due to particle-hole symmetry, $T_\ua (\omega) = T_\da (-\omega)$.
(a,b) Broadened curves at $B / T_\mr{K} = 10^2$ and $10^3$, plotted versus (a) $\omega/B$ and (b) $\omega$,
with different broadening schemes (adaptive or conventional)
and number of $z$-shifts ($n_z = 4$ and $8$).
The blue and red crosshairs in (a) indicate the value and uncertainty for
the peak positions and heights extrapolated to the limit $\alpha_z \rightarrow 0$ for adaptive and conventional schemes, respectively,
as derived from \Fig{fig5_extrap_Kondo}.
The extrapolated peak position at $B = 10^2 T_\mr{K}$ is consistent
with the DMRG result (dotted line, taken from Ref.~\onlinecite{Weichselbaum2009}).
(c) Discrete spectral data $\bar{A}/\omega$ for the data in panel (a) for the parameters as indicated.
}
\label{fig4_largeB_Kondo}
\end{figure}

\begin{figure}
\centerline{\includegraphics[width=.49\textwidth]{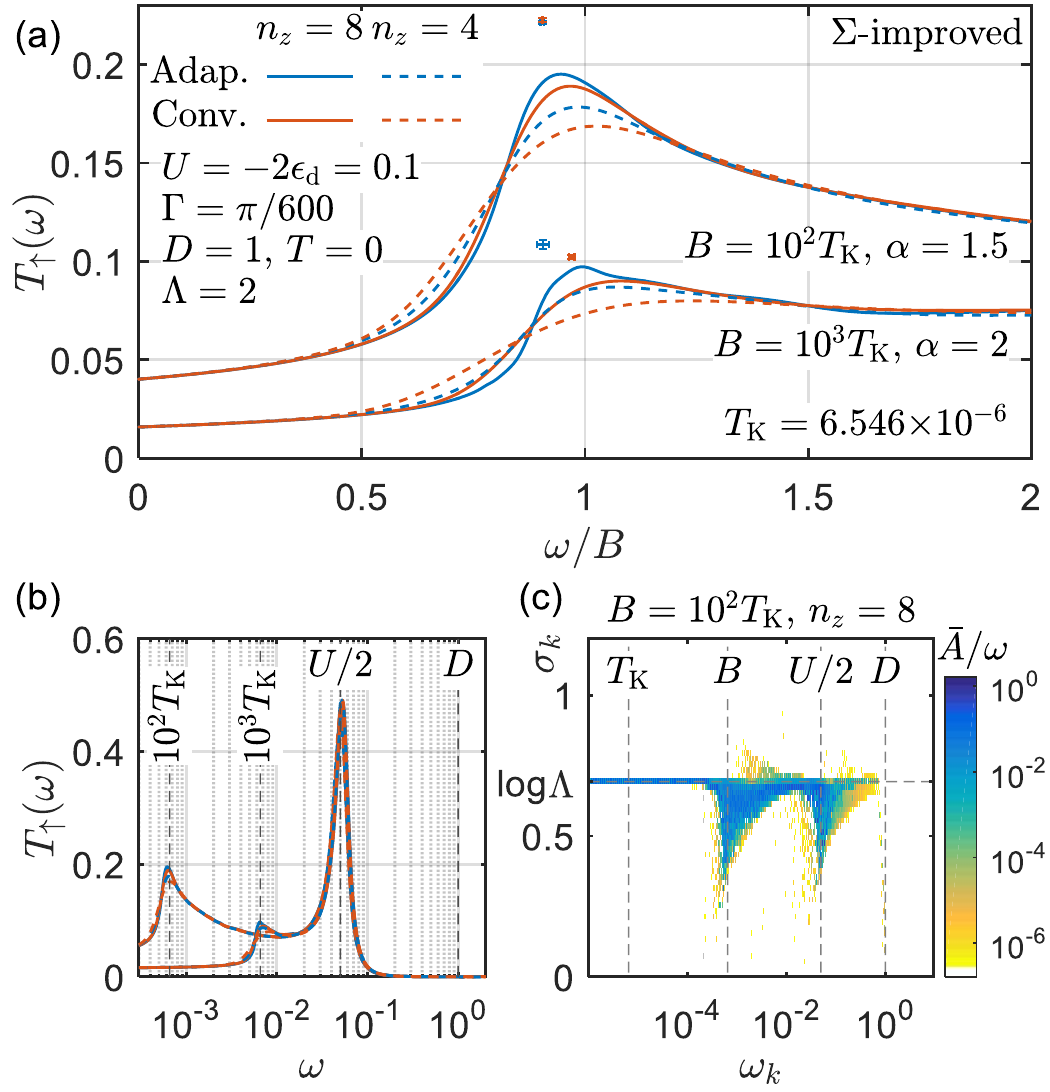}}
\caption{
$T$-matrix of the Anderson model for large magnetic field $B \gg T_\mr{K}$ at $T=0$.
Due to particle-hole symmetry, $T_\ua (\omega) = T_\da (-\omega)$.
(a,b) Broadend curves at $B / T_\mr{K} = 10^2$ and $10^3$, versus (a) $\omega/B$ and (b) $\omega$,
with different broadening schemes (adaptive or conventional)
and number of $z$-shifts ($n_z = 4$ and $8$).
All curves are improved by utilizing the self-energy $\Sigma$.
The blue and red crosshairs in (a) indicate value and uncertainty of
the peak positions and heights extrapolated to the limit $\alpha_z \rightarrow 0$
for adaptive and conventional schemes, respectively, as extracted from \Fig{fig5_extrap_Anderson}.
(c) Discrete spectral data $\bar{A}/\omega$ for the data in panel (a) for the parameters as indicated.
}
\label{fig4_largeB_Anderson}
\end{figure}

\subsection{Kondo and SIAM at large $B$}
\label{sec_largeB}

Next we apply a large magnetic field $B \gg T_\mr{K}$, which splits the Kondo peak of $T (\omega) = \tfrac{1}{2} \sum_{\nu = \ua,\da} T_\nu (\omega)$
in between the different spins $\nu$.
The split Kondo peaks are located at $\omega = \pm(B - \Delta \omega_B)$ for $\nu = \ua$ and $\nu = \da$, respectively.
The shift $\Delta \omega_B \simeq - B / [2 \log (B/T_\mr{K})]$ is suggested by analytic RG calculations~\cite{Rosch2003,Garst2005} and numerically confirmed by density-matrix renormalization group results~\cite{Weichselbaum2009}.
Here by having $B\neq 0$, the symmetry $\mr{SU}(2)_{\mr{spin}} \otimes \mr{SU}(2)_{\mr{ph}}$
above is reduced to $\mr{U}(1)_{\mr{spin}} \otimes \mr{SU}(2)_{\mr{ph}}$,
where we kept up to 2,000 multiplets in each step of iterative diagonalization.

As illustrated in Figs.~\ref{fig4_largeB_Kondo} and \ref{fig4_largeB_Anderson}, the adaptive broadening systematically improves the resolution of
the Kondo peaks for $B \gg T_\mr{K}$, in addition to the sharp edge at $|\omega| = D$ in the Kondo model and the Hubbard side peak at $|\omega| = U/2$ in the SIAM.
Overall we observe that for a fixed ratio of $B / T_\mr{K} \gg 1$, the position $\omega_{\max}$ of the Kondo peak is slightly larger in the Kondo model
(see also Figs.~\ref{fig5_extrap_Kondo} and \ref{fig5_extrap_Anderson} below).
The overall line shape of the curves are similar otherwise.
As $B$ increases, the peak line shapes remain qualitatively similar as a function of $\omega / B $,
except that the peak position slightly shifts
towards $\omega / B = 1$ and that the peak height is reduced.
Eventually, for the SIAM when $B \gtrsim U/2$, the Kondo peak
merges with the Hubbard side peak.

For the SIAM as well as the Kondo model, we choose a slightly larger $\alpha = 2$ for $B = 10^3 T_\mr{K}$ than $\alpha = 1.5$ for $B = 10^2 T_\mr{K}$
to smear out the residual noise from $z$-averaging at $n_z=8$ [cf.~\Sec{sec:z-limits}]. This noise persists even if we adjust the width of the outmost discretization interval in different ways as a function of $z$, or if we increase the number of kept states.
Overall we again observe that the adaptive broadening
leads to improved performance.

\begin{figure}
\centerline{\includegraphics[width=.49\textwidth]{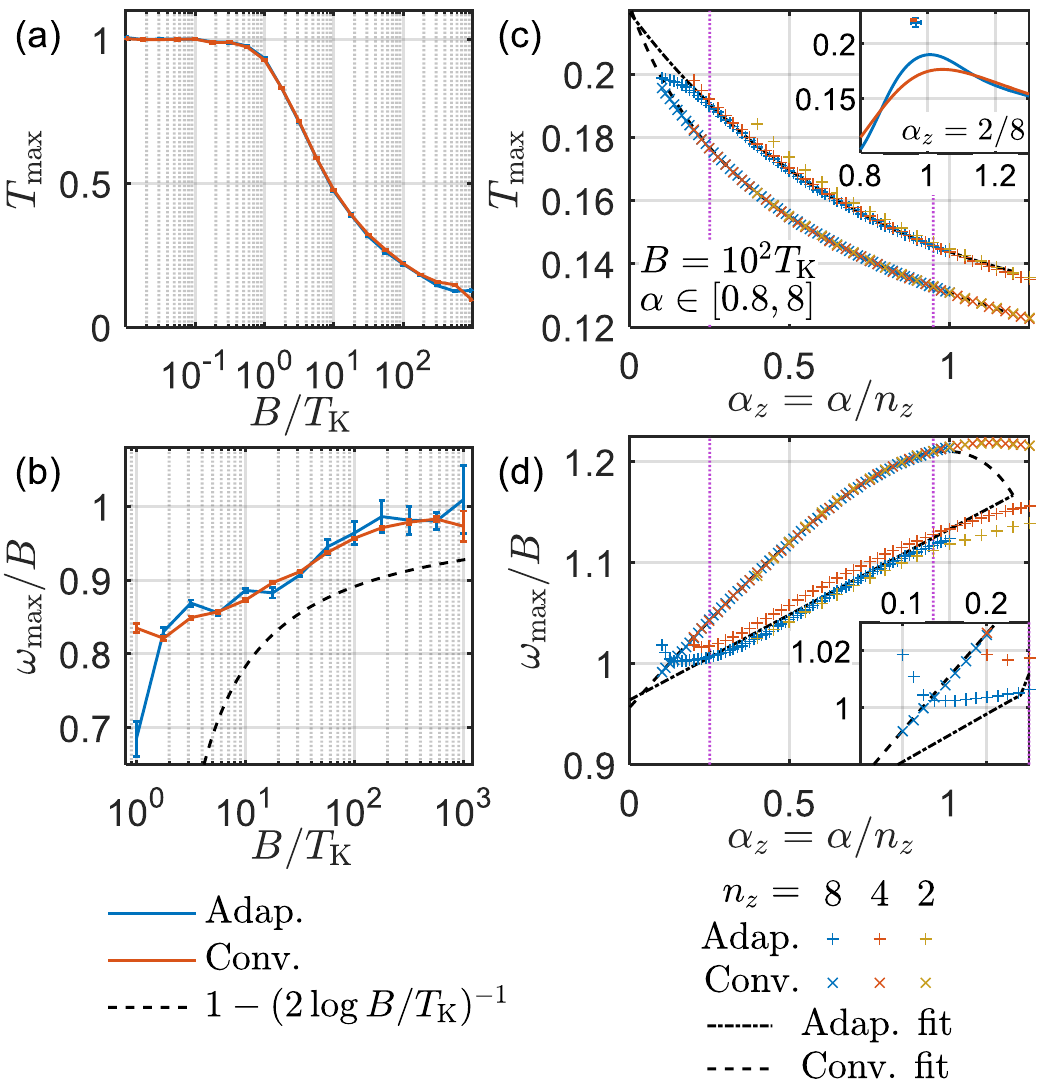}}
\caption{
Extrapolation of the Kondo peak in $T_\ua (\omega)$
for the Kondo model towards the ``continuum limit'' $\alpha_z \equiv \alpha / n_z \to 0$,
in terms of (a) heights $T_{\max}$ and (b) positions $\omega_{\max}$ vs.~$B/T_\mr{K}$.
For $\omega_{\max}$, we compare with the analytic prediction (dashed line). 
At $B=100\,T_\mr{K}$, we show (c) $T_{\max}$ and (d) $\omega_{\max}$ vs.~$\alpha_z$ (see Fig.~\ref{fig4_largeB_Kondo} for remaining parameters).
For each magnetic field, we extrapolate $T_{\max}$ and $\omega_{\max}$ to $\alpha_z \to 0$
by fitting the data points from $n_z = 4, 8$ within the fitting range
$[0.25, 0.95]$ as indicated by the vertical dotted lines with 
a Pad{\'e} approximant of order [2/1] (dashed and dash-dot lines).
The inset in (c) shows the adaptively (blue) and the conventionally (red) broadened curves of $T_\ua (\omega)$
at $B = 100T_\mr{K}$, $\alpha = 2$ and $n_z = 8$, 
where $\alpha_z = 0.25$ is on the lower edge of the fitting range. 
The extrapolated peak positions and heights are depicted as crosshairs.
The inset in (d) shows a close-up of the data points and the fitted curves of $\omega_{\max}/B$ vs.~$\alpha_z$ at small $\alpha_z$.
The upturn of $\omega_{\max}$ for the smallest $\alpha_z$
appears with the onset of underbroadening.
}
\label{fig5_extrap_Kondo}
\end{figure}

\begin{figure}
\centerline{\includegraphics[width=.49\textwidth]{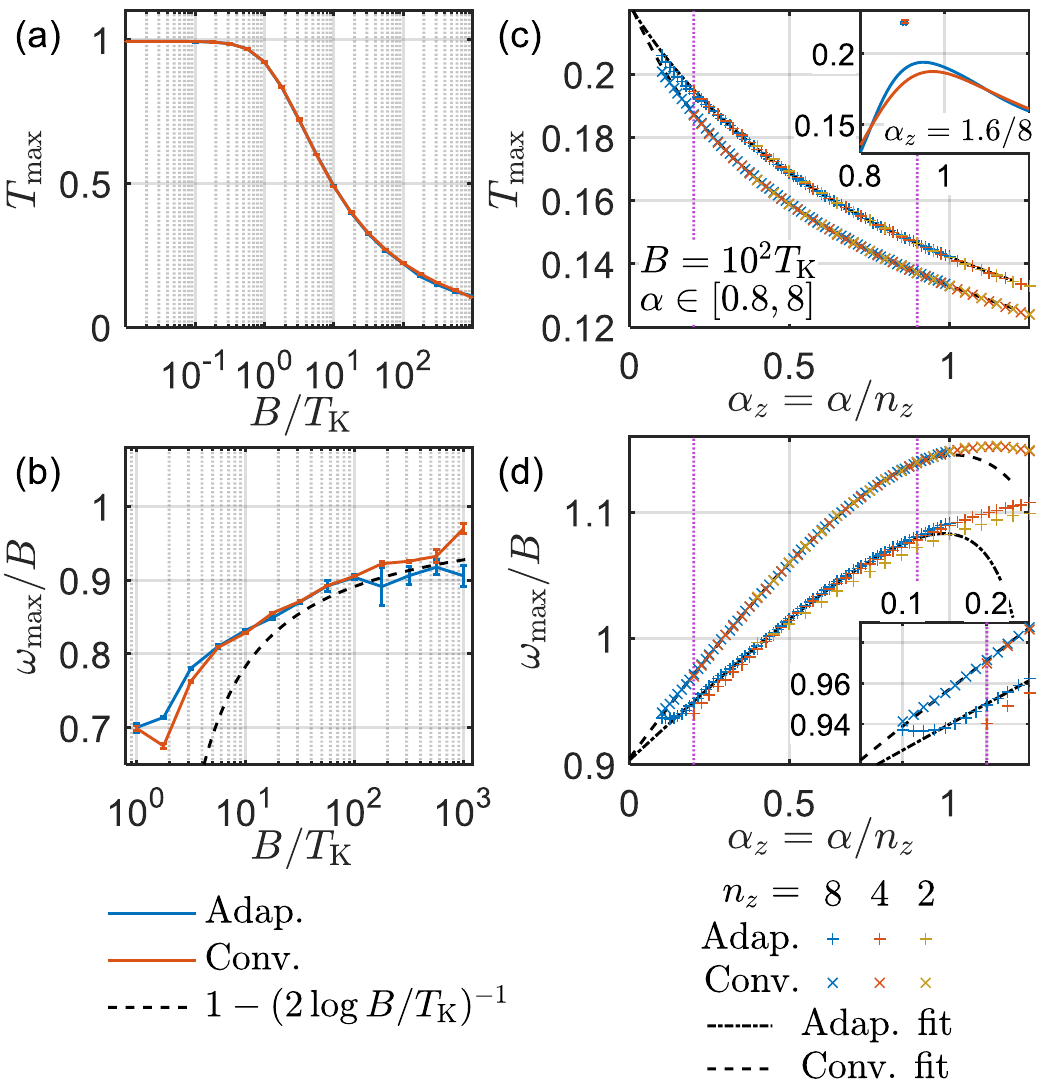}}
\caption{
Same as \Fig{fig5_extrap_Kondo}, but for the SIAM.
The fitting range in (c,d) is $\alpha_z \in [0.2, 0.9]$.
}
\label{fig5_extrap_Anderson}
\end{figure}

We estimate the height $T_{\max}$ and the position $\omega_{\max}$
of the Kondo peaks in the continuum limit,
by extrapolating their values at finite broadening $\bar{\sigma}_k \propto \alpha_z \equiv \alpha / n_z$ to $\alpha_z \rightarrow 0$.
In Fig.~\ref{fig5_extrap_Kondo}(c,d) and Fig.~\ref{fig5_extrap_Anderson}(c,d), 
we plot the heights $T_{\max}$ and positions $\omega_{\max}$ of the peak vs.~$\alpha_z$ 
obtained from different choices of $n_z$ and $\alpha$.
For both, adaptive as well as conventional broadening,
$T_{\max}$ and $\omega_{\max}$ show consistent dependency solely on the ratio $\alpha_z$.
In detail, the consistency is somewhat better within the
conventional approach (since uniformly overbroadened),
whereas minor offsets between different
$n_z$ values can be observed in the adaptive broadening.

For the extrapolation $\alpha_z \to 0$
we introduce lower and the upper cutoffs $\alpha_z^{\mr{min}}$ and $\alpha_z^{\mr{max}}$, respectively,
for the range of $\alpha_z$ to be included in the extrapolation.
Here $\alpha_z^{\mr{min}}$ is required to avoid
resolving the underlying discrete frequencies due to finite
$\Lambda$ and $n_z$
[see the insets of Figs.~\ref{fig5_extrap_Kondo}(d) and~\ref{fig5_extrap_Anderson}(d)].
Very large $\alpha_z$, on the
other hand, leads to excessive overbroadening such
that peak height and position become dependent on
the line shape of the spectral data over a wider region,
thus invalidating simple lower-order polynomial fits.
For example, in Figs.~\ref{fig5_extrap_Kondo}(d) and \ref{fig5_extrap_Anderson}(d)
we observe a qualitative change in the extracted peak position
$\omega_{\max}$ for $\alpha_z \gtrsim 1$ which we attribute
to excessive overbroadening.
For the results in this Section, we use the fitting range
$\alpha_z \in [0.25, 0.95]$ for the Kondo model and
$\alpha_z \in [0.2, 0.9]$ for the SIAM.
For the fitting, we use a Pad\'{e} approximant of order [2/1], i.e.~the ratio of a quadratic over a linear polynomial.
We estimate the error bar for the extrapolated value for $\alpha_z \to 0$
as the $95\%$ confidence interval out of this fit.

The extrapolated values $T_{\max}$ and $\omega_{\max}$ vs.~$B / T_\mr{K}$ are presented in Fig.~\ref{fig5_extrap_Kondo}(a,b) for the Kondo model and Fig.~\ref{fig5_extrap_Anderson}(a,b) for the Anderson model.
$T_{\max}$ shows a crossover around $B \sim T_\mr{K}$, which smoothly connects the value at $B = 0$ (e.g.~see Figs.~\ref{fig3_B0_Kondo} and \ref{fig3_B0_Anderson}) to the regime of large $B$ (e.g.~see Figs.~\ref{fig4_largeB_Kondo} and \ref{fig4_largeB_Anderson}).
We also compare our NRG result of $\omega_{\max} / B$ with the analytic prediction\cite{Rosch2003,Garst2005}
$1 - \Delta \omega_B / B = 1 - (2 \log B / T_\mr{K})^{-1}$
[black dashed lines in Figs.~\ref{fig5_extrap_Kondo}(b) and \ref{fig5_extrap_Anderson}(b)].
While the extrapolated $\omega_{\max} / B$ for the Kondo model systematically deviates from this analytical prediction, 
the one for the SIAM traces the analytic prediction more closely.
Besides that for the SIAM we exploit self-energy to get improved spectral data,
the difference with the Kondo model may also result from the fact that
the Kondo model is affected by finite bandwidth $D$ whereas by having $U\ll D$ the SIAM is not (e.g.~see Ref.~\onlinecite{Hanl2014}).

Overall, for both models the adaptive broadening
clearly gives consistently better peak resolution in terms of $T_{\max}$ and $\omega_{\max}$ for any finite $\alpha_z$, 
as compared to the conventional broadening
[see Figs.~\ref{fig5_extrap_Kondo}(c,d) and \ref{fig5_extrap_Anderson}(c,d)].
The extrapolation to the ``continuum limit'',
i.e.~$\alpha_z \to 0$, works accurately for magnetic fields
up to $B=100 T_\mr{K}$ [see Figs.~\ref{fig4_largeB_Kondo}(a)
and \ref{fig4_largeB_Anderson}(a)]. In particular, it is consistent
across conventional or adaptive broadening, and in the
case of the Kondo model, one also sees excellent
agreement of the peak position with accurate
DMRG simulations [\Fig{fig4_largeB_Kondo}(a)
with data reproduced from Ref. \onlinecite{Weichselbaum2009}];
the difference in the remaining line shape
is attributed to the
drastically finer discretization grid employed in the DMRG simulations
as compared to the $\Lambda=2$ for the NRG data here.
For $B \gg100 T_\mr{K}$, such as $B =10^3\, T_\mr{K}$
in Figs.~\ref{fig4_largeB_Kondo}(a)
and \ref{fig4_largeB_Anderson}(a),
the extrapolation is less reliable leading to larger error bars. The reason for this
is that the line-shape becomes less peak shaped,
but more and more step-like which increases the
uncertainty in the determination of the position of the
peak-maximum.

\subsection{Larger $\Lambda$}
\label{sec_largeL}

\begin{figure}
\centerline{\includegraphics[width=.49\textwidth]{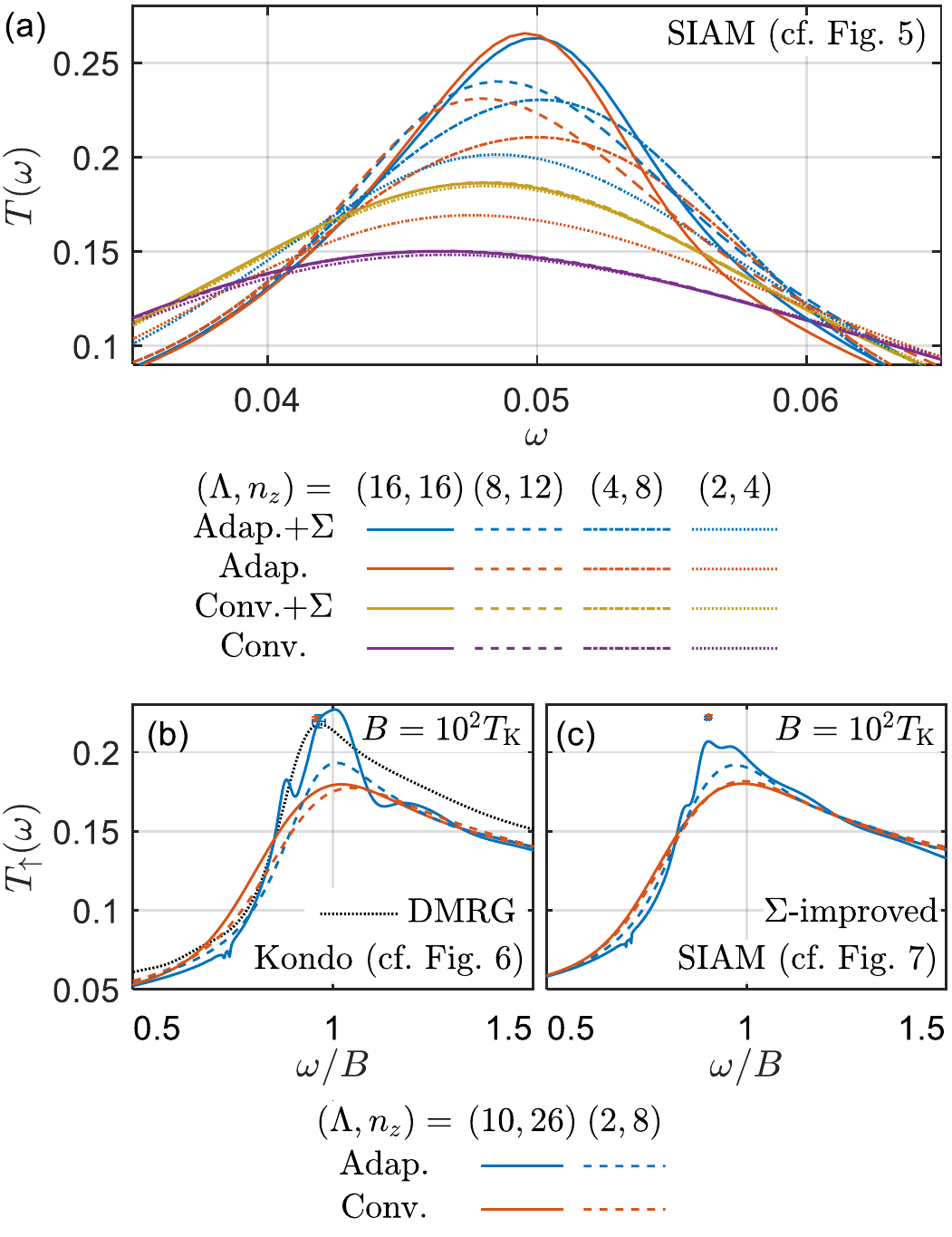}}
\caption{
Dependence of $T$-matrices on $\Lambda \in[2,16]$,
for the models discussed in Secs.~\ref{sec_B0}--\ref{sec_largeB} and Figs.~\ref{fig3_B0_Anderson}--\ref{fig4_largeB_Anderson}.
Here we use $\alpha=2$, and for comparable accuracy across
the wide range of $\Lambda$, we employ an energy-based truncation
with $E_{\rm{trunc}}=10$, throughout.
(a) Zoom into the Hubbard side peaks of the SIAM at $B=0$
(see \Fig{fig3_B0_Anderson} for system parameters).
Here we choose different pairs of $(\Lambda, n_z)$ with
a constant ratio $(\log \Lambda)/n_z$.
This implies the same spectral resolution within the conventional broadening scheme
in that the brown and purple lines (lower two rows in the legend)
(nearly) lie on top of each other.
Panels (b) and (c) show the $T$-matrix for 
(b) the Kondo and (c) the SIAM at large magnetic field ($B = 10^2 T_\mr{K}$; see Figs.~\ref{fig4_largeB_Kondo} and \ref{fig4_largeB_Anderson} for system parameters, respectively).
Here again $\log(\Lambda)/n_z$=$\log(10)/26 \simeq \log(2)/8$ was chosen comparable.
Note that by truncating with respect to energy ($E_{\mr{trunc}}=10$)
rather than fixed number of states,
the curves of $\Lambda = 2$ here slightly differ from the ones in Figs.~\ref{fig3_B0_Anderson}--\ref{fig4_largeB_Anderson}.
For comparison, we again also plot the extrapolated peak position and height (blue and red crosshairs) and the DMRG data (black dotted line), adapted from Figs.~\ref{fig4_largeB_Kondo}(a) and \ref{fig4_largeB_Anderson}(a).
}
\label{fig10_largeL}
\end{figure}

In the previous Sections, we have used constant $\Lambda = 2$,
a typical choice for NRG simulations of single-band impurity models,
and kept $N_\mr{keep} = 500$ or $2000$ multiplets to
demonstrate well-converged results.
For multi-band problems, however, this choice of $\Lambda = 2$ is no more practical
since the required number of multiplets $N_\mr{keep}$ increases
exponentially with the number of bands, and the computational
cost scales like $\mathcal{O}(N_\mr{keep}^3)$.
This problem can be partly ameliorated by using significantly larger $\Lambda$,
where the energy scales of the Wilson chain sites $\sim \Lambda^{-n/2}$ are better separated.
Then the entanglement in the ground state of the Wilson chain
is lowered\cite{Weichselbaum2012:mps}, with the effect that
a significantly smaller $N_\mr{keep}$ suffices for well-converged results.
Accordingly, $4 \lesssim \Lambda \lesssim 10$ is frequently used to simulate multi-band models accurately and feasibly.
For such larger $\Lambda$, $z$-averaging is absolutely crucial
to cancel out oscillatory behavior due to enhanced discretization artefacts.\cite{Bulla2008}

Here we demonstrate that, depending on the situation,
one can achieve nearly comparable spectral
resolution for $\Lambda$ as large as 16. 
This is further supported by the adaptive broadening 
where the resolution enhancement
is more significant for larger $\Lambda$.
For this, we revisit the systems considered in Secs.~\ref{sec_B0} and \ref{sec_largeB}.
In \Fig{fig10_largeL}(a), we depict the Hubbard side peaks of the
SIAM in the absence of magnetic field
for different choices of $(\Lambda, n_z)$ while keeping
$\log(\Lambda)/n_z$ constant to ensure comparable spectral
resolution. Indeed, the conventional broadening results in
hardly distinguishable curves (brown and purple lines).
However, when turning on the adaptive broadening, this further resolves
the Hubbard side peaks with increasing $\Lambda$. Specifically, for $\Lambda$ as large as
$\Lambda = 16$, (i) the curves are still smooth without
any discretization blips, and (ii) the
adaptively broadened curves already show
comparable peak shape with and without self-energy improvement.
This suggests that the peak height is converged for these curves!

In \Fig{fig10_largeL}(b) and (c), we show the split Kondo peaks by $B = 10^2 T_\mr{K}$ for the Kondo model and the SIAM, respectively.
By again choosing comparable $\log(\Lambda)/n_z$,
the conventionally broadened curves for different $\Lambda$ are 
nearly on top of each other.
In contrast, the adaptively broadened curves for larger $\Lambda = 10$ 
again show enhanced resolution, even though here for $B\gg T_\mr{K}$
at the price of more
pronounced discretization artefacts.
While this may not come
as a surprise -- after all we are using a hugely crude disretization
based on $\Lambda=10$ -- one can still observe that aside from
discretization related blips, the mean of the resulting curve still
moves around a consistent overall lineshape. In particular, the
data in \Fig{fig10_largeL}(b-c)
is still consistent with the $\alpha\to0$ extrapolated peak height
(symbols) or the DMRG data [black dotted line in panel (b)] replicated from
Figs.~\ref{fig4_largeB_Kondo} and \ref{fig4_largeB_Anderson}.
Also in \Fig{fig10_largeL}(b), the height of the plateau
for $\omega>B$ is consistent within the NRG \emph{in the entire
range} $\Lambda \in [2,10]$. Therefore we believe the NRG
data is more reliable for $\omega>B$ than the replicated DMRG
data which itself may have suffered from inaccuracies in a
numerically necessarily unstable deconvolution scheme.

We conclude this Section with a few technical remarks.
In order to compare the data for different $\Lambda$
at the same footing with similar accuracy,
the state space trunction within the NRG\cite{Bulla2008}
proceeds along an energy-based threshold $E_\mr{trunc} = 10$ with adaptively varying number of kept multiplets $N_\mr{keep}$,
rather than a fixed $N_\mr{keep}$, using the conventions adopted in Ref.~\onlinecite{Weichselbaum2011},
keeping all states without truncation for the first five NRG iterations.
For example, in Figs.~\ref{fig10_largeL} (b-c) where we have used $\mr{U}(1)_\mr{spin} \otimes \mr{SU}(2)_\mr{ph}$ symmetry,
the resulting number of kept states (multiplets) is
$\lesssim 6700$ ($2700$) for $\Lambda = 2$, and $\lesssim 300$ ($150$) for $\Lambda = 10$, respectively.
This demonstrates the clear reduction in computational cost by using larger $\Lambda$.

\section{Results at finite $T$}
\label{sec_finT}

\begin{figure}
\centerline{\includegraphics[width=.49\textwidth]{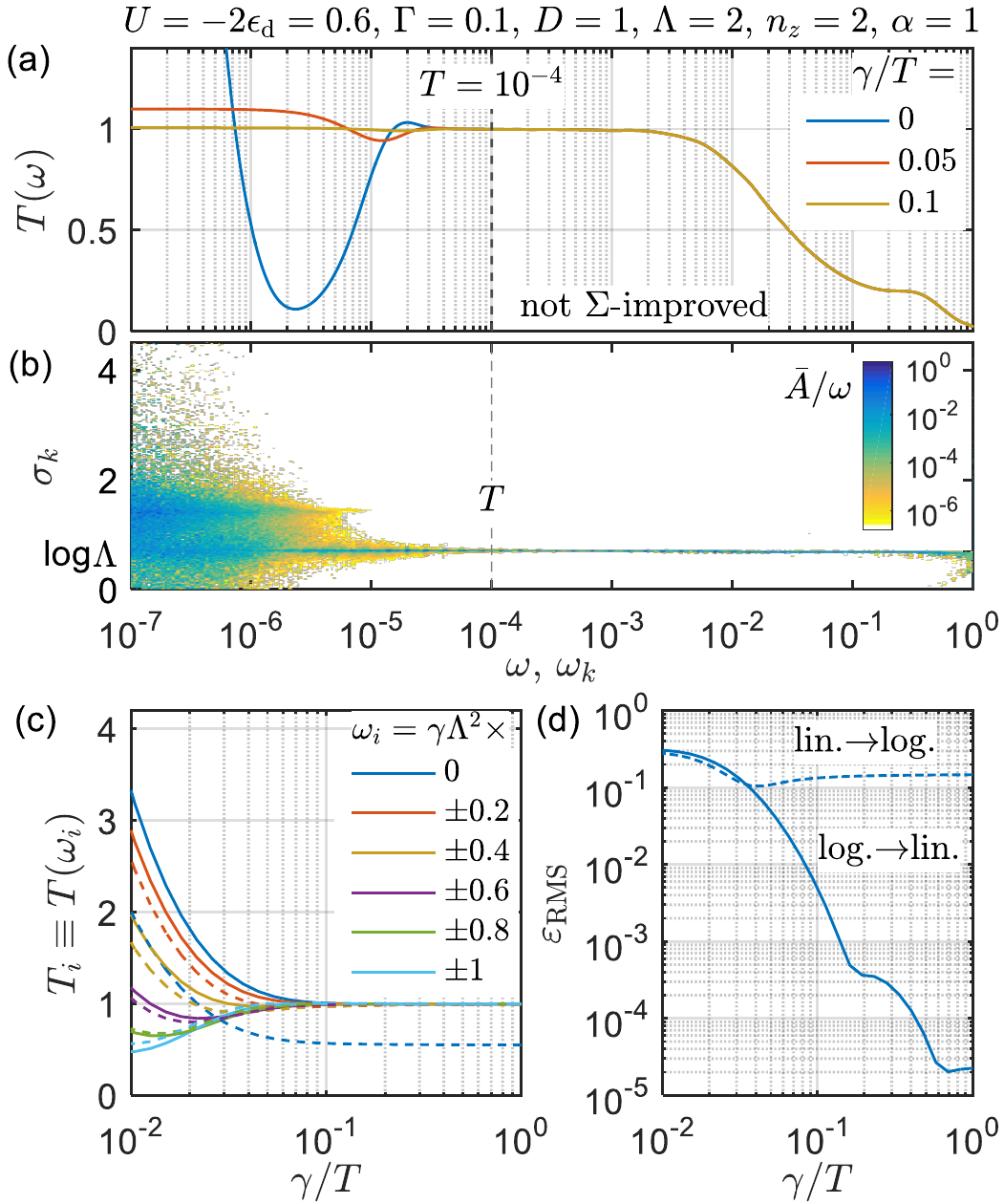}}
\caption{
$T$-matrix of the SIAM at finite $T \ll T_\mr{K}$.
(a) Broadened curves for different $\gamma$ and (b) binned discrete data $\bar{A} / \omega$ for $T (\omega)$.
Contrary to the confined distribution of $\bar{A} / \omega$ at $T \lesssim \omega_k \lesssim T_\mr{K}$ and $\sigma_k \simeq \log \Lambda$,
the weights at $|\omega | \ll T$ spread widely along $\sigma_k$.
This leads to distinct irregular behavior in $T(\omega)$
at $|\omega | \ll T$ after the logarithmic Gaussian broadening [blue solid line in (a)].
The secondary convolution with width $\gamma \lesssim T$ smears out sharp fluctuations without the blip artefact, keeping features at higher $\omega$ untouched [red and brown solid lines in (a)].
(c) Spectral function values $T_i \equiv T (\omega_i)$ at frequencies
$\omega_i = \gamma \Lambda^2 \times \{-1, -0.8, -0.6, \cdots, 1\}$ vs.~$\gamma/T$ (see text).
(d) Error $\varepsilon_\mr{RMS}$ of quadratic polynomial fit vs.~$\gamma/T$ [see \Eq{eq:eps_rms}].
In (c,d), solid lines are obtained by applying first the log-Gaussian and then the linear broadening kernels, as in \Eq{A:cont}.
In comparison, dashed lines are obtained by the opposite order of broadening;
the order of broadening kernels is essential to obtain smooth curves at $|\omega| \lesssim T$ [cf.~\Sec{sec:Tfinite}].
}
\label{fig6_finT}
\end{figure}

\begin{figure}
\centerline{\includegraphics[width=.49\textwidth]{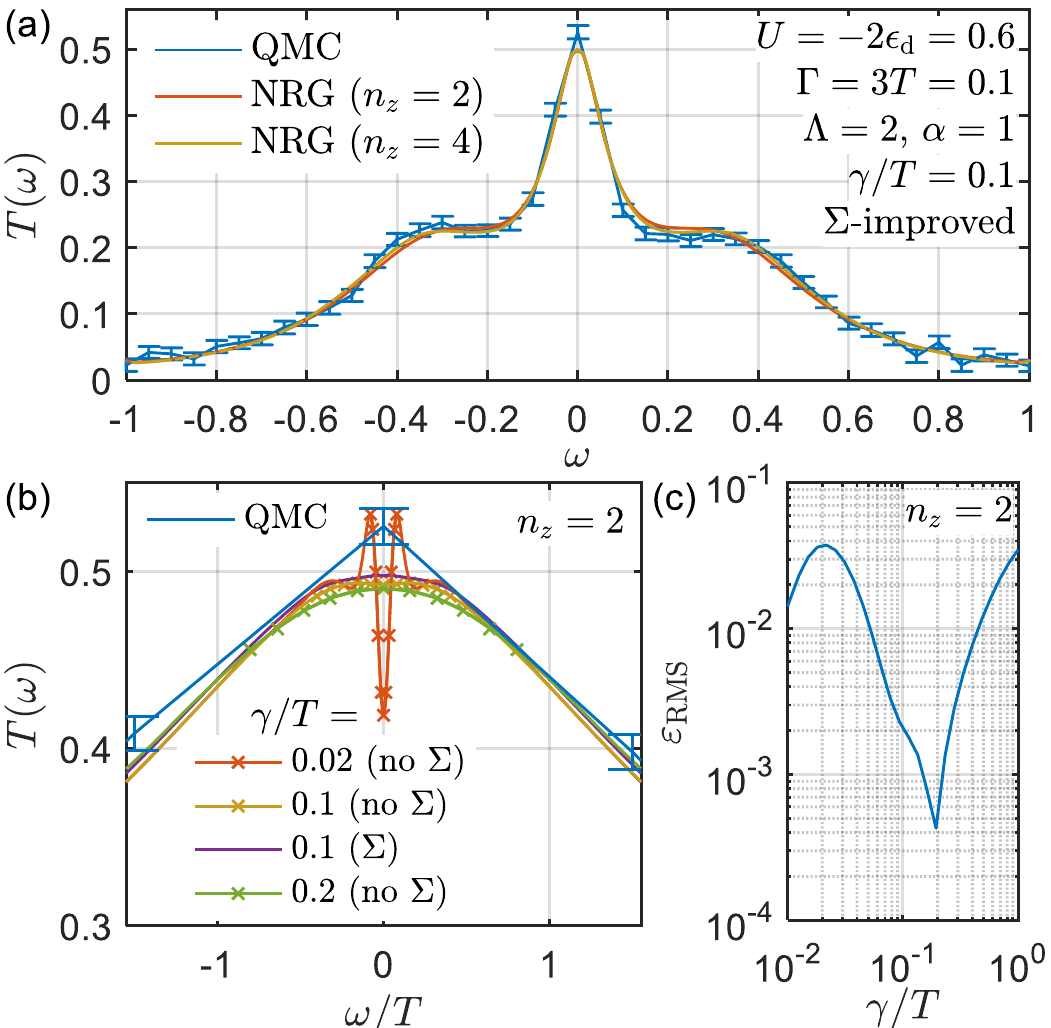}}
\caption{
$T$-matrix of the
SIAM at $T = D/30 \simeq 2T_\mr{K}$.
(a) Comparison of NRG and quantum Monte Carlo results (data from Ref.~\onlinecite{Cohen2014})
(b) Low-frequency region of $T(\omega)$ for different $\gamma$, 
where $\times$ symbols are the discrete points $T_i$ at $\omega_i = \gamma \Lambda^2 \times \{-1, -0.8, -0.6, \cdots, 1\}$ vs.~$\gamma/T$ to estimate
the smoothness of the curve at $|\omega | \sim \gamma$.
(c) Error $\varepsilon_\mr{RMS}$ of quadratic polynomial fit for discrete points
vs.~$\gamma / T$ [cf.~\Eq{eq:eps_rms}].
}
\label{fig7_QMC}
\end{figure}

In the previous Section we analyzed frequencies $|\omega| \gg T=0^+$
where log-Gaussian broadening is well-suited. For frequencies $|\omega|\lesssim T$,
however, this needs to be amended by a linear broadening scheme,
as discussed in \Sec{sec_broad} earlier.
In this Section, we present the results for the SIAM at three different
temperature scales $T \ll T_\mr{K}$, $T \sim T_\mr{K}$, and $T \gg T_\mr{K}$,
as shown in Figs.~\ref{fig6_finT}-\ref{fig9_largeT}, respectively.
This is followed by a comparison to QMC data at intermedate
temperatures, as well as a discussion on a systematic way to determine the optimal value of linear broadening width $\gamma$.

In Fig.~\ref{fig6_finT}, we depict $T (\omega)$ at finite $T \ll T_\mr{K}$,
where the Kondo peak height is close to the perfect transmission $T(0) \simeq 1$.
Using log-Gaussian broadening only [blue solid line in Fig.~\ref{fig6_finT}(a)]
the spectral curve shows irregular behavior at $| \omega | \ll T$ due to
effectively finite chain length induced by finite $T$ [cf.~\Sec{sec:Tfinite}].
Note though, that the log-Gaussian broadened spectral data is already smooth for frequencies $\omega \ge T$.
After further convolving the curve with the linear kernel $\delta_\gamma$ in Eq.~\eqref{delta:lin},
a smooth curve emerges for $\gamma = T/10$ (still very small as compared to temperature $T$).

At elevated temperatures $T \gtrsim T_\mr{K}$,
the Kondo physics becomes suppressed by thermal fluctuations.
\Fig{fig7_QMC}(a) shows that the Kondo peak height at $T \simeq 2 T_\mr{K}$ is almost halved from the value at $T = 0^+$.
At even higher temperature $T \simeq 100 T_\mr{K}$,
the Kondo physics is fully suppressed, leaving behind only the two Hubbard
side peaks, as illustrated in \Fig{fig9_largeT}.

We observe that the discrete data $\bar{A} / \omega$ shows
a pronounced spread along $\sigma_k$ as $|\omega_k|$ decreases below $T$:
while the spread appears only at $|\omega_k| \lesssim T/3$ for $T \ll T_\mr{K}$ [\Fig{fig6_finT}(b)],
the spread becomes even more pronounced over all $|\omega_k| \leq D$ for $T \gg T_\mr{K}$ [\Fig{fig9_largeT}(b)].
Given an interacting model, this spread in broadening width $\sigma_k$ naturally
tends to smear out spectral data on the energy scale of temperature.

Next we analyze how the low-frequency region of $T(\omega)$ changes with $\gamma$.
This is illustrated in
Figs.~\ref{fig7_QMC}(b) and \ref{fig9_largeT}(d)
for $T \sim T_\mr{K}$ and for $T \gg T_\mr{K}$, respectively.
For both cases, $\gamma /T = 0.1$ sufficiently smooths the curve,
and at the same time minimizes overbroadening (brown and purple lines).
In contrast, the curves of $\gamma / T < 0.1$ show discretization-related oscillations
[red line in \Fig{fig7_QMC}(b), red and blue lines in \Fig{fig9_largeT}(d)].
Meanwhile, for $\gamma / T > 0.1$, the curve segments over the
interval $|\omega| < T/2$ are overall shifted
(relative to the curves for $\gamma / T \leq 0.1$)
hence indicating the onset of overbroadening (green lines).

\subsubsection{Comparison to QMC data}

\begin{figure}
\centerline{\includegraphics[width=.49\textwidth]{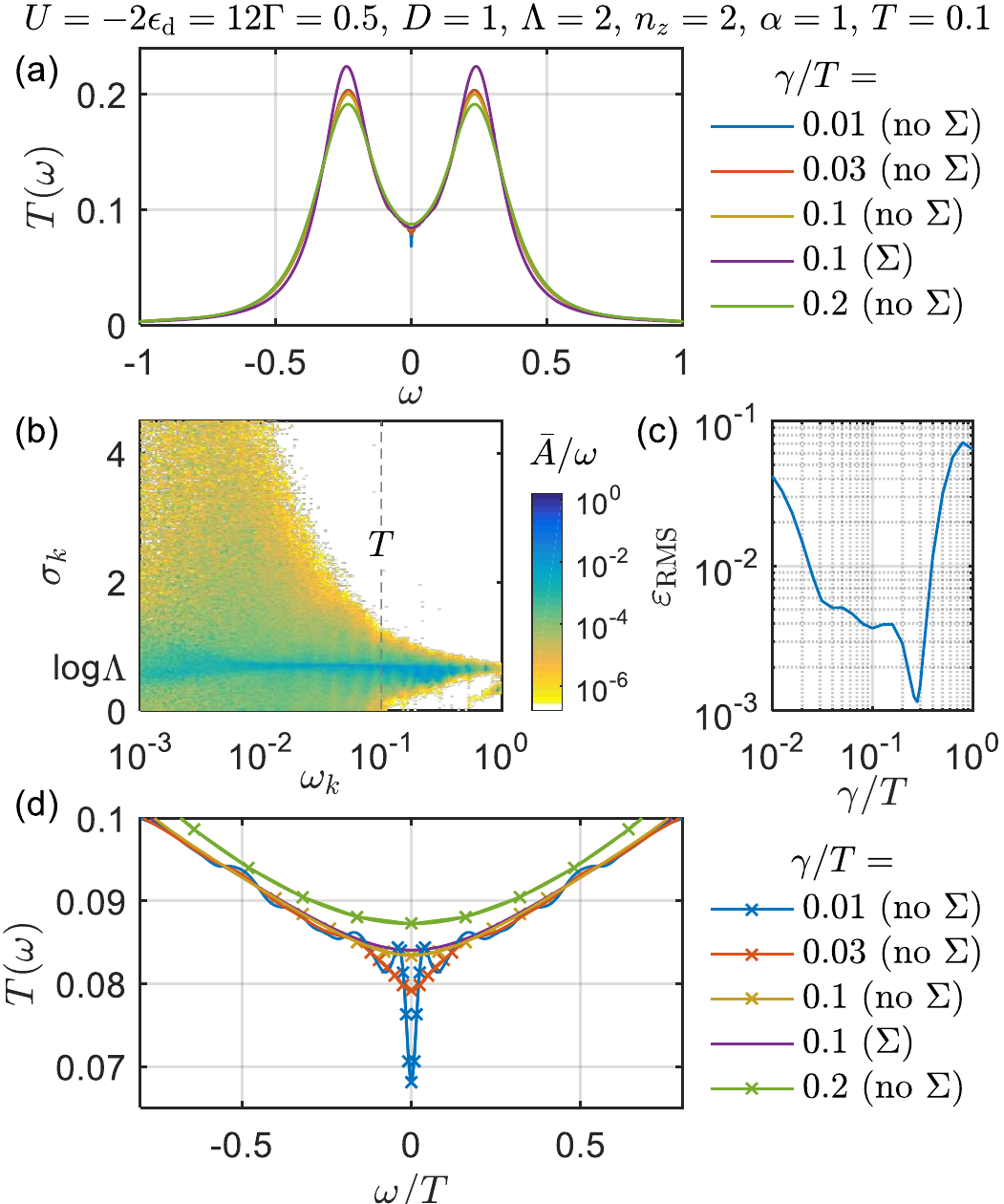}}
\caption{
$T$-matrix of the SIAM at large temperature ($T \simeq 100 T_\mr{K}$).
(a) Broadened curves for various $\gamma$.
Due to high temperature, the Kondo peak is completely suppressed while the Hubbard side peaks persist at $\omega = \pm U/2$.
Using the self-energy ($\Sigma$) improves the spectral resolution of the Hubbard peaks.
The $\Sigma$-improved curve for $n_z = 2$ at $\gamma / T = 0.1$ (purple line) shows a height of the Hubbard peaks
that is comparable with a non-$\Sigma$-improved curve for $n_z = 8$ and the same $\gamma / T = 0.1$ (not shown).
(b) Binned discrete data $\bar{A}/\omega$ for $T (\omega)$.
While the weights at $|\omega_k| < T$ spread widely along $\sigma_k$ for $T \ll T_\mr{K}$ [see \Fig{fig6_finT}(b)],
here a certain spread also appears at $|\omega_k| \gtrsim T$.
(c) Same as \Fig{fig7_QMC}(c).
(d) Same as \Fig{fig7_QMC}(b).
}
\label{fig9_largeT}
\end{figure}

At intermediate temperatures $T \sim T_\mr{K}$,
we can compare our NRG result with recent state-of-the-art
quantum Monte Carlo (QMC) calculation. \cite{Cohen2014}
The results are presented in Fig.~\ref{fig7_QMC}(a) at $T = D / 30 \simeq 2 T_\mr{K}$.
Though our NRG result mostly lies within the error bar of the QMC result,
at $\omega = 0$ the Kondo peak height of the NRG is systematically about 5\% lower, 
and thus clearly outside the QMC error bar.
NRG, however, is known to produce consistent accurate results
at $\omega=0$ to within 1\% at arbitrary temperature
[e.g.~see the perfect consistency with Friedel sum-rule at $T\ll T_\mr{K}$ in \Fig{fig6_finT}(a)].
We speculate that the QMC may have overestimated the peak height or underestimated the error bar.
Also the current setting of $U = -2\epsilon_\mr{d}$ yields the particle-hole symmetry,
which appears in the $T$-matrix as $T(\omega) = T (-\omega)$.
While the NRG accurately reproduces the symmetry, the QMC data is only barely particle-hole symmetric within its error bars.
For the results in Figs.~\ref{fig6_finT}-\ref{fig9_largeT},
we kept up to 500 multiplets (about 3300 states) exploiting SU(2) spin and SU(2) particle-hole symmetry in each step of the iterative diagonalization.
For a single $z$-shift, the calculation of the entire spectral data took about ten minutes on an 4-core workstation.
Note that in the QMC calculation the band edges of $\Gamma (\epsilon)$
are slightly smoothened, but this smoothing does not affect the NRG result
as we explicitly checked.

\subsubsection{Optimal choice for linear broadening width $\gamma$}
\label{sec_gamma}

The width $\gamma$ of the linear broadening that follows
the log-Gaussian broadening, so far,
is a parameter that needs to be tuned by hand.
While sufficiently large $\gamma \lesssim T$ ensures
a smooth final spectral curve, $\gamma=T$ typically 
already results in clear overbroadening.
Ideally, $\gamma$ is chosen as small as possible to just smear out 
discretization related irregularities at $|\omega| \ll T$ that are out of reach for the
log-Gaussian broadening,  but keeps the overbroadening
of already smooth physical features at $|\omega|\gtrsim T$
to a minimum. 
The optimal choice for the linear broadening width $\gamma$
for the spectral data for $|\omega|<T$ necessarily depends
on the underlying physics.
Hence a systematic determination of $\gamma$ is desirable.

Foremost, this requires a measure to quantify ``smoothness''
of the data $|\omega|\lesssim T$ on a linear scale
including its transition to the log-Gaussian broadening.
We start from spectral data that has already been
broadened in the \emph{entire} frequency range by an adaptive
log-Gaussian scheme as described earlier.
Then we estimate the smoothness of the curve obtained
after the linear broadening with any $\gamma \leq T$:
we sample $T(\omega)$ on a linear grid of frequencies $\omega_i$ in the close vicinity of $\omega = 0$,
and check whether the sampled values $T(\omega_i)$ can be well fit by a quadratic functions.
If so, $T(\omega)$ is regarded as smooth.
This procedure avoids applying
the linear broadening to the entire spectral range.
Hence the determination of an optimal $\gamma$ as outlined
here, is numerically cheap.

To be specific, we consider (i) a \emph{linear} frequency
range $\omega \in [-\omega_l, \omega_l]$ with $\omega_l
\equiv \gamma \Lambda^2$.
(The subscript $l$ stands for linear.)
Eventually, we will demand 
that the broadened resulting spectral data closely follows
a quadratic polynomial in this interval.
By including the scale factor $\Lambda^{n_l}$ with $n_l=2$
in the definition of $\omega_l$,
the linear broadening width $\gamma$ is extended to at least
$n_l$ intervals of the underlying logarithmic discretization for
both positive and negative frequencies. This ensures that
the analysis of the resulting smoothness clearly reaches across
into the log-Gaussian broadened frequency range.
(ii) Next we split this frequency range $\omega \in [-\omega_l, \omega_l]$
into a uniformly spaced linear grid $\omega_i$ of $m_l$ frequencies
$\{\omega_i\} = [-\omega_l, \ldots, \omega_l]$, where we choose
$m_l = 5n_l+1=11$. By being odd, this includes the frequency
$\omega=0$, and by being $> 4n_l$, this ensures sufficient
resolution into the underlying logarithmic discretization grid
towards the log-Gaussian broadened data at $|\omega| \sim \omega_l$.
(iii) We then compute the linearly broadened data
$\{T_i\} \equiv T(\{\omega_i\};\gamma)$ 
at the above linear grid of $m_l$ frequencies only [see \Fig{fig6_finT}(c)], and (iv) perform
a quadratic fit on this data $(\omega_i, T_i)$, 
which results
in a quadratic polynomial $\tilde{T}(\omega)$ for the data range $|\omega|\le \omega_l$.
We define the error estimate of the fit as the
normalized distance of the data to this quadratic fit,
\begin{eqnarray}
      \varepsilon_{\mr{RMS}} \equiv \frac{ \sqrt{\langle \delta^2 T_i \rangle} }{\langle T_i \rangle} ,
\label{eq:eps_rms}
\end{eqnarray}
where $ \langle T_i \rangle \equiv \tfrac{1}{m_l} \sum_{i=1}^{m_l}  T_i $ 
is the average
and $\la \delta^2 T_i \ra \equiv \tfrac{1}{m_l - 3} \sum_{i=1}^{m_l} (T_i - \tilde{T}(\omega_i))^2$
is the standard error of fitting
[note that $m_l - 3$ is the statistical degree of freedom, as there are three coefficients of quadratic polynomial fitting function].
(v) The error $\varepsilon_\mr{RMS}$
provides the desired estimate for the smoothness of the fully
broadened data for frequencies $|\omega| \lesssim T$
that now includes both, the initial log-Gaussian,
as well as the subsequent linear broadening of width $\gamma$.
A small value of $\varepsilon_\mr{RMS}$ implies that $T(\omega)$ indeed behaves quadratically within the fitting range $[-\omega_l, \omega_l]$,
indicating that discretization artefacts have been satisfactorily smoothed out.

In Figs.~\ref{fig6_finT}(d), \ref{fig7_QMC}(c), and \ref{fig9_largeT}(c), we show the dependence of the fitting error estimate $\varepsilon_\mr{RMS}$ vs.~$\gamma$
in three different regimes $T \ll T_\mr{K}$, $T \sim T_\mr{K}$, and $T \gg T_\mr{K}$, respectively.
At $T \ll T_\mr{K}$,
the error $\varepsilon_{\mr{RMS}}$ monotonously decreases with
increasing $\gamma$ [solid line in \Fig{fig6_finT}(d)], due to the featureless flat behavior of $T(\omega)$ 
at $|\omega| \lesssim T \ll T_\mr{K}$.
As seen in \Fig{fig6_finT}(a),
the spectral data is visibly smooth for $\gamma/T = 0.1$, which corresponds to $\varepsilon_{\mr{RMS}} \simeq 0.002$.

The situation becomes different at higher temperatures,
in that $T(\omega)$ actually has structure, e.g.~curvature at $|\omega|\lesssim T$.
Therefore the strong rise of $\varepsilon_{\mr{RMS}}$ towards larger $\gamma$
at $\gamma / T > 0.2$ [\Fig{fig7_QMC}(c)] and at $\gamma / T > 0.3$ [\Fig{fig9_largeT}(c)] indicates the onset of overbroadening:
given that the line shape within the linear frequency window $|\omega| \leq T$
is no longer exactly quadratic [see Figs.~\ref{fig7_QMC}(b) and \ref{fig9_largeT}(d)],
the fitting quality of a quadratic fit in the range $[-\omega_l, \omega_l]$ (with $\omega_l = \gamma \Lambda^2$)
necessarily deteriorates with larger $\gamma$.

Conversely, the strong increase of $\varepsilon_{\mr{RMS}}$
towards small $\gamma$ in \Fig{fig7_QMC}(c) [as already also seen in
\Fig{fig6_finT}(d)] indicates the onset of discretization artefacts due to
underbroadening.
In \Fig{fig7_QMC}(c) this occurs for $\gamma / T < 0.2$.
With $\gamma / T = 0.2$ still visibly overbroadened, though,
due to the smallness of $\varepsilon_{\mr{RMS}}$
at $\gamma / T = 0.2$, $\gamma$ can be further reduced
as long as discretization artefacts are still weak.
Similar to \Fig{fig6_finT}(d) where we estimated
$\varepsilon_{\mr{RMS}} \simeq 0.002$ for smooth data,
the same requirement also leads to $\gamma/T = 0.1$ here.

A similar picture emerges for large temperatures as seen in
\Fig{fig9_largeT}(c): a strong increase in $\varepsilon_\mr{RMS}$,
for $\gamma / T > 0.3$ due to overbroadening, and for $\gamma / T < 0.03 $
due to underbroadening. In the large temperature case, however,
there emerges an intermediate window for $\gamma / T \in [0.03, 0.3]$
that shows more irregular behavior of $\varepsilon_\mr{RMS}$ yet at
small values. This is interpreted as a consequence of the interplay
of the underlying discrete data with the broadening as well as the
intrinsic line shape of the spectral function.

Based on the above observations, we therefore suggest the following
procedure to determine the optimally minimal $\gamma$:
(i) Obtain the dependence $\varepsilon_\mr{RMS}$ vs.~$\gamma$,
over a wide range on a log-scale, e.g.~$\gamma/T \in [0.01,1]$.
Since the evaluation of $\varepsilon_\mr{RMS}$ does not require the linear broadening over all frequencies,
this step can be done efficiently.
By analyzing its dependence, we can identify the regimes
where under- or over-broadening occurs.
(ii) If only the underbroadening behavior appears [e.g.~$T \ll T_\mr{K}$ as in \Fig{fig6_finT}(d)],
choose the value of $\gamma / T$ at which $\varepsilon_\mr{RMS}$ passes through a certain threshold, e.g.~$0.002$.
So we have chosen $\gamma / T = 0.1$ in \Fig{fig6_finT}(d).
(iii) If the underbroadening behavior appears directly next to the overbroadening behavior [e.g.~$T \sim T_\mr{K}$ as in \Fig{fig7_QMC}(c)],
there will be two values of $\gamma / T$ at which $\varepsilon_\mr{RMS}$ passes through the threshold, e.g.~$0.002$.
We choose the smaller $\gamma$, since the larger one clearly overbroadens the curve;
thus $\gamma / T = 0.1$ is chosen in \Fig{fig6_finT}(d) as well.
(iv) If there exists a more irregular region between the under- and over-broadening regimes [e.g.~$T \gg T_\mr{K}$
as in \Fig{fig9_largeT}(d)],
take the geometric mean (or the midpoint on a log-scale, equivalently)
of the smallest overbroadening $\gamma / T$ and the largest underbroadening $\gamma / T$ to stay equally far from either side.
For the example of \Fig{fig9_largeT}(d), these two values are $\gamma / T \simeq 0.03$
and $0.3$, respectively. Incidentally, this again results in an optimal
$\gamma / T \simeq 0.1$ at slightly enhanced $\varepsilon_\mr{RMS} \simeq 0.003$.
As seen in \Fig{fig9_largeT}(a,d), this trades off the shift of the curve segment due to overbroadening against the oscillation due to underbroadening.

\section{Summary}
\label{sec_conc}

We have developed an adaptive scheme which broadens each discrete spectral weight 
individually based on its position's sensitivity on $z$-shifts
in the underlying logarithmic discretization. For frequencies
$|\omega| \lesssim T$ we have developed a systematic scheme to keep the required linear broadening of
width $\gamma$ to a minimum.

The additional computational cost for the adaptive broadening is 
minor: (i) Only the matrix elements of the perturbation $\mr{d}H / \mr{d}z$
need to be computed to estimate the broadening width.
(ii) The discrete spectral data then is collected on a 2-dimensional array, i.e.~with dimensions frequency $\omega_k$ and broadening $\sigma_k$.
(iii) The actual broadening
is part of the post-analysis of the fdm-NRG run. Its cost is linear
proportional to the number $n_\sigma$  of bins in the broadening
$\sigma_k$, where in practice a linear grid of $n_\sigma \simeq 50$ 
bins should suffice for a linear range
$\sigma \in [0, 2] \times \log\Lambda$.

The adaptive broadening presented here systematically
improves spectral resolution.
With increasing $n_z$,
it converges much more quickly to the analytically known
line shapes of non-interacting models than the conventional broadening.
Yet also for interacting models, the adaptive broadening converges faster to the ``continuum limit''.
In the limit of infinite $z$-averaging, i.e.~$n_z \to \infty$,
the adaptive approach necessarily coincides with the 
conventional approach. However, the infinite $n_z$ limit
is not accessible within the NRG (and moreover is biased
by the existence of the band edges, see \Sec{sec:z-limits}).
The adaptive scheme presented in this paper
systematically improves spectral resolution from dynamical NRG data at finite $n_z \ge 2$.
Therefore the proposed adaptive broadening should benefit
two widely used applications of the NRG:
(i) DMFT calculations in the quest to deal with structured
bath hybridization functions and, quite generally,
(ii) multi-band (effective) impurity calculations
to obtain maximal spectral resolution at necessarily
larger coarse graining in energy.

\section{Acknowledgement}

We thank to Jan von Delft for fruitful discussion, Emanuel Gull for providing us with his QMC data, and Theo Costi for useful feedback.
S.~L.~acknowledges support from the Alexander von Humboldt Foundation and the Carl Friedrich von Siemens Foundation.
A.~W.~was supported by DFG (Nanosystems Initiative Munich, and Grant No.~WE4819/2-1).

\appendix

\section{Logarithmic discretization}
\label{app_LD}

In this appendix we discuss the numerical calculation of the derivatives of hopping amplitudes $\mr{d} t_{n} / \mr{d} z$ and of onsite energies
$\mr{d} \varepsilon_{n} / \mr{d}z$.
This analysis proceeds independently for each flavor $\nu$.
Hence, for simplicity, we skip the flavor index
in the following discussion.

For simplicity, we also focus on the case that the impurity and the bath are coupled
via the quadratic hybridization $H_\mr{cpl}$
with the hybridization function $\Gamma(\omega)$
as in the SIAM in \Eq{H_cpl}, 
but our argument is easily
extendible to the Kondo model.
As usual within the NRG, we define the 
logarithmic discretization intervals symmetrically for
positive and negative energies, i.e.~$I_{k,+}^{(z)}
\equiv (  \epsilon_{k+1}^{(z)},  \epsilon_{k}^{(z)} ]$
and $I_{k,-}^{(z)}
\equiv [ -\epsilon_{k}^{(z)},  -\epsilon_{k+1}^{(z)} )$,
with
\begin{equation}
  \epsilon_{k}^{(z)} = \begin{cases}
  D , & k = 0, \\
  D \Lambda^{-k+1-z} , & k > 0.
  \end{cases}
\label{LDint}
\end{equation}
In the process of coarse-graining,
we replace the bath continuum within each interval
$I_{ks}^{(z)}$ with $k=0,1,2,\ldots$ and $s\in\{+,-\}$ by a single discrete level at energy $\xi_{ks}^{(z)}$
which couples to the impurity with amplitude $t_{ks}^{(z)} = (\tfrac{\gamma_{ks}^{(z)}}{\pi})^{1/2}$.
The total hybridization of the discrete levels
represents the hybridization of the continuum
still, since 
\begin{eqnarray}
   \int \Gamma (\epsilon) \, \mr{d}\epsilon
= \sum_{k s} \underbrace{
\int_{I_{k s}^{(z)}} \Gamma (\epsilon) \, \mr{d}\epsilon
}_{\equiv \gamma_{k s}^{(z)}} \ \text{,}
\label{gamma_nz}
\end{eqnarray}%
which defines $\gamma_{k s}^{(z)}$.The continuum limit may be restored
by $z$-averaging over $n_z\rightarrow \infty$ $z$-shifts
that are uniformly distributed within $z\in (0,1]$,
i.e.~$\tfrac{1}{n_z} \sum_z \to \int_0^1 \mr{d}z$.
With focus on the non-interacting bath only,
the energy $\xi_{k s}^{(z)}$ then needs to
satisfy \cite{Zitko2009}
\begin{equation}
\Gamma ( \xi_{k s}^{(z)} ) = 
\gamma_{k s} ^{(z)}
\left/ \left| \tfrac{\mr{d}\xi_{k s}^{(z)}}{\mr{d}z} \right| \right. ,
\label{ZitkoEq1}
\end{equation}
which is a differential equation in the continuous variable
$x \equiv k+z \in [0, \infty]$.

Via coarse-graining in energy space,
the continuous star Hamiltonian
$H_\mr{cpl} + H_\mr{bath}$ in Eq.~\eqref{totHam}
becomes the discrete star Hamiltonian $H_\mr{star} = \psi_{\mr{star}}^{(z)\dagger} \mb{H}_{\mr{star} }^{(z)} \psi_{\mr{star}}^{(z)}$
where $ \psi_\mr{star}^{(z)} \equiv (
d, a_{0+}^{(z)}, a_{1+}^{(z)}, \cdots, a_{0-}^{(z)},
a_{1-}^{(z)}, \cdots )^\mr{T} $
with $a_{k s}^{(z)}$ the annihilation operator of the discretized bath level $(k s)$ for a given $z$, and
\begin{equation}
\begin{gathered}
\begin{aligned}
&\mb{H}_{\mr{star},\nu}^{(z)} = 
\left(
\begin{array}{c | c c c | c c c}
0 & t_{0+}^{(z)} & t_{1+}^{(z)} & \cdots & t_{0-}^{(z)} & t_{1-}^{(z)} & \cdots \\
\hline
t_{0+}^{(z)} & \xi_{0+}^{(z)} & & & & & \\
t_{1+}^{(z)} & & \xi_{1+}^{(z)} & & & & \\
\vdots       & & & \ddots & & & \\
\hline
t_{0-}^{(z)} & & & & \xi_{0-}^{(z)} & & \\
t_{1-}^{(z)} & & & & & \xi_{1-}^{(z)} & \\
\vdots       & & & & & & \ddots   
\end{array} \right).
\end{aligned}
\end{gathered}
\label{starHam}
\end{equation}
Here ``star'' means a star-shaped tree geometry of how the impurity and the bath levels are coupled;
the impurity couples to the bath levels and there is no direct coupling between the bath levels.
Then by Lanczos tridiagonalization with the starting vector
$[1,0,0,\ldots]^\mr{T}$, i.e.~leaving the impurity level $d$ as it is, the star Hamiltonian in Eq.~\eqref{starHam} can be exactly mapped onto a chain geometry,
$H_{\mr{chain}}^{(z)} \equiv  \psi_{\mr{chain}}^{(z)\dagger} \,
\mb{T}^{(z)} 
\, \psi_{\mr{chain}}^{(z)}$, with
$\psi_{\mr{chain}}^{(z)} \equiv (
d, f_{0}^{(z)}, f_{1}^{(z)}, f_{2}^{(z)}, \cdots)^\mr{T}$ and the tridiagonal Hamiltonian matrix
\begin{eqnarray}
\mb{T}^{(z)}
&=& [\mb{U}^{(z)}]^\dagger \, \mb{H}_{\mr{star}}^{(z)} \, \mb{U}^{(z)} \\
&=& \begin{pmatrix}
0 & t_{0}^{(z)} & & & \\
t_{0}^{(z)} & \varepsilon_{0}^{(z)} & t_{1}^{(z)} & & \\
 & t_{1}^{(z)} & \varepsilon_{1}^{(z)} & t_{2}^{(z)} & \\
 & & t_{2}^{(z)} & \varepsilon_{2}^{(z)} & \ddots  \\
 & & & \ddots & \ddots \\
\end{pmatrix}.
\label{chainHam2}
\end{eqnarray}%
By construction, the discrete intervals $(k s)$ depend on
the underlying $z$-shift. Therefore also
$\psi_\mr{star}^{(z)}$ and subsequently
also $\psi_\mr{chain}^{(z)}$ refer to a $z$-dependent
coarse-grained basis set. From the point of view of the
impurity in the NRG, however, the set $\{f_n\}$ simply
refers to a set of one-particle states. In this sense,
the $z$-dependence in $\psi_\mr{chain}^{(z)}$ is irrelevant
and can be ignored.

Therefore by only considering the $z$-dependence of
the tridiagonal matrix $\mb{T}^{(z)}$,
the perturbation $\mr{d}H/\mr{d}z$
translates into $\mr{d}\mb{T} /\mr{d}z$.
Based on the defining equations (\ref{gamma_nz}--\ref{chainHam2}),
we therefore simply compute
\begin{eqnarray}
\frac{\mr{d}\mb{T}^{(z)}}{\mr{d} z} \simeq 
\frac{1}{\delta z} \left(
   \mb{T}^{(z + \delta z)} - \mb{T}^{(z)}
\right)
\end{eqnarray}
This perturbation simply consists of the numerical derivatives
$\mr{d}\varepsilon_{n}/\mr{d}z$ and $\mr{d} t_{n}/\mr{d}z$
which enter the diagonal and first off-diagonal in
$\mr{d}\mb{T}/\mr{d}z$, respectively.
These, however, are numerical derivatives that need to be based on two different Lanczos
tridiagonalizations with slightly offset $z$-shifts.
In practice, we chose $\delta z=0.01/n_z$.
Note that the perturbation $\mr{d} \mb{T} /\mr{d}z$ clearly
cannot be simply related to the differentiation of 
$\mr{d}\xi_{k s}/\mr{d}z$ and
$\mr{d} t_{k s}/\mr{d}z$ in the star geometry
since the unitary transformation $\mb{U}^{(z)}$
itself is $z$-dependent.

For any hybridization function $\Gamma(\omega)$ that
is finite at $\omega=0$, the asymptotic behavior for large $n$
for the hoping amplitudes is $t_{n}^{(z)} \propto
1/\Lambda^{{\tfrac{n}{2}}+z}$ and therefore
$(\mr{d} t_n^{(z)} / \mr{d}z) / t_n^{(z)}  = \mr{d}\log(t_n^{(z)})/\mr{d}z \simeq
- \log\Lambda$ [eventually, this needs to
be multiplied by the global factor $1/n_z$ to get the full
perturbation; cf.~\Eq{sigma:k}].
The onsite energies $\varepsilon_n^{(z)}$ have nontrivial
asymptotic behavior that decays at least as $\Lambda^{-n}$,
unless the particle-hole symmetry enforces $\varepsilon_n^{(z)} = 0$.

\section{Equation \eqref{sigmaij} in case of degeneracy}
\label{app_deg}

For the estimate of the broadening width of discrete
spectral data based on its sensitivity on $z$-shifts,
we made use of the Hellmann-Feynmann theorem
in \Eq{sigmaij} in the main text.
Here we address the implications of (accidental)
degeneracy in the energy eigenstates.

Degeneracy of eigenstates typically occurs
due to symmetry such as
particle number conservation, particle-hole symmetry,
total spin conservation, etc.
In practical NRG calculations, 
these symmetries are fully exploited to 
strongly reduce numerical cost in terms of
memory consumption and CPU time. \cite{Weichselbaum2012:sym}
Accidental degeneracy can be neglected,
since this always may be removed by an infinitesimal 
external perturbation, which in a numerical setting
may be interpreted as numerical noise that always weakly
lifts exact accidental degeneracy anyway.

Therefore degenerate eigenstates typically arise due to symmetry. 
As such they are (i) part of the same multiplet if the full symmetry setting includes non-abelian symmetries
(e.g.~degenerate states within a given symmetry
multiplet, say, of some total spin $S$) or
(ii) distinguishable by different quantum numbers
(such as spin-component $S_z$)
if a reduced symmetry setting is used for the
simulation itself.
In either case, the matrix elements of $\mr{d}H/\mr{d}z$
will be block-diagonal with respect to symmetry by the Wigner-Eckart theorem,
since the perturbation $\mr{d}H/\mr{d}z$ relates to a 
scalar Hamiltonian (note that $z$-shifts do not break 
the symmetry of the original Hamiltonian).

Consequently the application of the Hellmann-Feynmann 
is legitimate, since degenerate eigenstates in different
symmetry sectors are distinguishable, i.e.~they do 
not mix. Conversely, 
degeneracy within a given
symmetry multiplet space has always a diagonal 
matrix representation, since the Clebsch-Gordan 
coefficients out of the Wigner-Eckart theorem for
a scalar operator are always proportional 
to an identity matrix. Hence the perturbation
will not mix in between different states
of the same multiplet since symmetry is preserved.

Overall, therefore this justifies that we can use the energy eigenstates $\ket{E_i}$ directly
obtained from the iterative diagonalization in Eq.~\eqref{sigmaij} without having to worry about degenerate subspaces.


\begin{thebibliography}{99}

\bibitem{Wilson1975} K. G. Wilson, Rev. Mod. Phys. {\bf 47}, 773 (1975).
\bibitem{Bulla2008} R. Bulla, T. A. Costi, and T. Pruschke, Rev. Mod. Phys. {\bf 80}, 395 (2008).
\bibitem{Georges1996} A. Georges, G. Kotliar, W. Krauth, and M. J. Rozenberg, Rev. Mod. Phys. {\bf 68}, 13 (1996).
\bibitem{Bulla1999} R. Bulla, Phys. Rev. Lett. {\bf 83}, 136 (1999).
\bibitem{Stadler2015} K. M. Stadler, Z. P. Yin, J. von Delft, G. Kotliar, and A. Weichselbaum, Phys. Rev. Lett. {\bf 115}, 136401 (2015).
\bibitem{Weichselbaum2007} A. Weichselbaum and J. von Delft, Phys. Rev. Lett. {\bf 99}, 076402 (2007).
\bibitem{Bulla2001} R. Bulla, T. A. Costi, and D. Vollhardt, Phys. Rev. B {\bf 64}, 045103 (2001).
\bibitem{Yoshida1990} M. Yoshida, M. A. Whitaker, and L. N. Oliveira, Phys. Rev. B {\bf 41}, 9403 (1990).
\bibitem{Campo2005} V. L. Campo and L. N. Oliveira, Phys. Rev. B {\bf 72}, 104432 (2005).
\bibitem{Zitko2009} R. \v{Z}itko and T. Pruschke, Phys. Rev. B {\bf 79}, 085106 (2009).
\bibitem{Freyn2009} A. Freyn and S. Florens, Phys. Rev. B {\bf 79}, 121102 (2009).
\bibitem{Anders2005} F. B. Anders and A. Schiller, Phys. Rev. Lett. {\bf 95}, 196801 (2005).
\bibitem{Anders2006} F. B. Anders and A. Schiller, Phys. Rev. B {\bf 74}, 245113 (2006).
\bibitem{Weichselbaum2012:mps} A. Weichselbaum, Phys. Rev. B {\bf 86}, 245124 (2012).
\bibitem{Costi2000} T. A. Costi, Phys. Rev. Lett. {\bf 85}, 1504 (2000).
\bibitem{Bulla1998} R. Bulla, A. C. Hewson, and T. Pruschke, J. Phys.: Condens. Matter {\bf 10}, 8365 (1998).
\bibitem{Weichselbaum2011} A. Weichselbaum, Phys. Rev. B {\bf 84}, 125130 (2011).
\bibitem{Stadler2013} K. M. Stadler, Master's thesis, Ludwig-Maximilians-Universit\"{a}t M\"{u}nchen (2013).
\bibitem{Osolin2013} \v{Z}. Osolin and R. \v{Z}itko, Phys. Rev. B {\bf 87}, 245135 (2013).
\bibitem{Hanl2014} M. Hanl and A. Weichselbaum, Phys. Rev. B {\bf 89}, 075130 (2014).
\bibitem{Weichselbaum2012:sym} A. Weichselbaum, Ann. Phys. {\bf 327}, 2972 (2012).
\bibitem{Weichselbaum2009} A. Weichselbaum, F. Verstraete, U. Schollw\"{o}ck, J. I. Cirac, and J. von Delft, Phys. Rev. B {\bf 80}, 165117 (2009).
\bibitem{Rosch2003} A. Rosch, T. A. Costi, J. Paaske, and P. W\"{o}lfle, Phys. Rev. B {\bf 68}, 014430 (2003).
\bibitem{Garst2005} M. Garst, P. W\"{o}lfle, L. Borda, J. von Delft, and L. Glazman, Phys. Rev. B {\bf 72}, 205125 (2005).
\bibitem{Cohen2014} G. Cohen, E. Gull, D. R. Reichman, and A. J. Millis, Phys. Rev. Lett. {\bf 112}, 146802 (2014).

\end{thebibliography}
\end{document}